%% file: main.tex
\documentclass[
    aip,
    jcp,
    amsmath,
    amssymb,
    reprint,
    numerical
]{revtex4-2}

\usepackage{mathtools}
\usepackage{physics}
\usepackage{bm}

\usepackage{siunitx}

\usepackage{graphicx}
\usepackage{booktabs}
\usepackage[dvipsnames]{xcolor}

\graphicspath{{figures/}}

\usepackage{microtype}

\usepackage{hyperref}

\hypersetup{
    colorlinks=true,
    linkcolor=MidnightBlue,
    citecolor=MidnightBlue,
    urlcolor=MidnightBlue
}

\usepackage[nameinlink,noabbrev]{cleveref}

\begin{document}
\title{
PULSUS: Direct Frequency-Domain Multidimensional Spectroscopy
Beyond the Impulsive Rotating-Wave Approximation
}

\author{Alfonso Castillo-Gonzalez}
\affiliation{
Department of Physics,
University of Houston,
Houston, Texas 77204-5005, USA
}
\email{ancasti4@CougarNet.UH.EDU}

\author{Eric R. Bittner}
\affiliation{
Department of Physics,
University of Houston,
Houston, Texas 77204-5005, USA
}
\affiliation{Institut Courtois, Universit\'e de Montr\'eal, 1375~Avenue~Th\'er\`ese-La\-voie-Roux, Montréal, Qu\'ebec H2V~0B3, Canada}
\email{ebittner@central.uh.edu}

\date{\today}
\input{Abstract}
\maketitle

\input{Introduction}

\input{Theory}

\input{Software}

\input{Validation}

\input{RegimeStudy}

\input{Conclusion}

\input{EndMatter}

\section*{REFERENCES}
\bibliography{references}

\end{document}

%% file: Abstract.tex
\begin{abstract}
We introduce PULSUS, a direct frequency-domain framework for
multidimensional spectroscopy of open quantum systems beyond the
impulsive rotating-wave approximation. Coherence-frequency intervals
are represented through Liouvillian resolvents, while finite Gaussian
pulses are incorporated through operator-valued functions of the
field-free Liouvillian. This enables direct evaluation on Cartesian
grids or arbitrary spectral coordinates while retaining molecular
evolution during the pulse envelopes and, when required, the complete
light--matter interaction.

The implementation is validated against analytical and independent
time-domain benchmarks and supports targeted spectral sampling that
reduces the number of evaluated coordinates when the response is
spectrally localized. A controlled hierarchy of finite-pulse,
rotating-wave, short-pulse, and impulsive models separates the errors
introduced by each approximation. For a coupled electronic dimer, the
total impulsive-RWA error is nonmonotonic in pulse duration: an
intermediate regime provides the highest accuracy, whereas ultrashort
pulses expose coupling-enabled beyond-RWA pathways and longer pulses
develop substantial finite-pulse corrections. These results show that
the impulsive and rotating-wave approximations cannot be assumed to
improve simultaneously and should be tested when quantitatively
interpreting multidimensional spectra. PULSUS provides a direct
frequency-domain framework for carrying out that comparison.
\end{abstract}

%% file: Introduction.tex
\section{Introduction}
\label{sec:introduction}

Multidimensional coherent spectroscopy provides direct access to
correlations between excitation and detection frequencies and has
become an important tool for resolving electronic couplings,
coherences, relaxation pathways, and many-body interactions in
molecular and condensed-phase systems
\cite{
Abramavicius2009CoherentMultidimensional,
Oliver2018MultidimensionalReview,
Hybl1998TwoDimensional}.
Interpreting these spectra generally requires comparison with a
microscopic dynamical model, making efficient numerical calculation of
nonlinear optical response functions an important part of the
spectroscopic analysis.

Theoretical treatments are often simplified by assuming impulsive
optical interactions and applying the rotating-wave approximation
(RWA).
These approximations provide a compact pathway picture and are useful
over broad regions of parameter space, but finite pulse duration can
introduce effects that are absent from the impulsive description.
Finite-pulse treatments have demonstrated modifications of spectral
amplitudes, phases, line shapes, and observed dynamical signatures,
with the importance of these effects depending on pulse bandwidth,
resonance conditions, and the molecular dynamics occurring during the
interaction
\cite{
Smallwood2017FinitePulse,
Perlik2017FinitePulse,
DoGelinTan2017FinitePulse,
Leng2017FinitePulseWidth}.
For sufficiently broadband fields, the spectral suppression of
counter-rotating light--matter processes can become less effective,
allowing non-RWA contributions to become appreciable
\cite{Smallwood2017FinitePulse,Tan2026RWA}.
Representative contemporary two-dimensional electronic spectroscopy
measurements employ ultrashort pulses in the $10$--$20~\mathrm{fs}$
range, including experiments using approximately $12~\mathrm{fs}$ and
$21~\mathrm{fs}$ pulses
\cite{Timmer2023Plasmon,RojasGatjens2023ManyExciton}.
Consequently, the validity of the impulsive and rotating-wave
approximations must ultimately be assessed relative to the molecular and
optical timescales of the system under study.

A variety of analytical and numerical strategies have been developed
for treating nonlinear spectroscopy outside the simplest impulsive
picture.
Mixed time--frequency formulations and finite-pulse response
constructions provide alternatives to direct multidimensional
time-domain propagation
\cite{
GelinDomcke2016AlternativeView,
DoGelinTan2017FinitePulse}.
Frequency-domain Green-function approaches have also been used to
describe nonlinear multi-exciton signals
\cite{Roslyak2013Quasiparticle}.
For open quantum systems, the UF$^2$ method evaluates nonlinear signals
through Fourier convolution in the eigenbasis of the Liouvillian and
can efficiently incorporate arbitrary sampled pulse envelopes
\cite{RoseKrich2021UF2}.
Automated diagram generation further provides a systematic treatment
of the additional temporal orderings that arise when pulses overlap
\cite{RoseKrich2021DiagramGenerator}.
Open-source packages such as QuDPy and QDT provide complementary
pathway-oriented and response-function frameworks for constructing
multidimensional optical signals
\cite{shah2023qudpy,Kenneweg2024QDT}.

The present work targets a complementary computational regime in which
the spectral coordinates themselves are the requested numerical
objects. Rather than constructing a multidimensional time-domain
interferogram and subsequently transforming it, we seek to evaluate
selected coherence frequencies directly while retaining molecular
evolution during finite Gaussian pulse envelopes. The objective is not
only to calculate a spectrum, but also to expose, within one
formulation, the consequences of successively introducing the
rotating-wave, short-pulse, and impulsive approximations.

Here we introduce a direct frequency-domain Liouville-space
framework that combines three complementary elements. First,
finite Gaussian pulses are incorporated through analytic
operator-valued functions of the field-free Liouvillian, retaining
molecular evolution during the pulse envelopes. Second, the
response is evaluated directly at Cartesian grids or arbitrary
paired spectral coordinates, without first constructing a uniform
delay-domain interferogram. Third, the complete molecular
interaction and its finite-pulse RWA, short-pulse RWA, and
impulsive-RWA reductions are placed within a single controlled
hierarchy, allowing the physical consequences of each approximation
to be isolated.

PULSUS is the Python implementation of this framework and supports
both first- and third-order response calculations. The present work
focuses on the third-order multidimensional formulation, where the
method combines analytic finite-pulse Liouvillian dressing, direct
spectral-coordinate evaluation, and controlled comparison of the
approximations entering the nonlinear response.

This organization is particularly useful when only selected regions
of a multidimensional spectrum are required, or when the validity of
the usual pulse and rotating-wave approximations is itself part of the
question. The present finite-pulse formulation treats finite-duration,
Gaussian pulses in the separated-pulse regime, where the chronological
ordering of the interactions remains well defined. Strongly overlapping
pulse sequences can introduce additional temporal orderings and response
pathways and therefore lie outside the present formulation.

Within this scope, PULSUS is complementary to finite-pulse convolution
and pathway-oriented approaches: it sacrifices arbitrary pulse-shape
and pulse-overlap generality in exchange for analytic finite-pulse
Liouvillian dressing and direct evaluation at the requested spectral
coordinates. The treatment remains perturbative in the optical field;
strong-field and nonperturbative two-dimensional spectroscopy constitute
a distinct regime that has been considered elsewhere~\cite{
AndaCole2021BeyondPerturbative,
BressanVanThor2021IntenseFields}.

We validate the finite-pulse implementation against the analytical
Gaussian-pulse benchmark of Smallwood \textit{et al.}, benchmark the
complete-interaction Model A independently against explicit
time-domain perturbative propagation, and verify agreement between
Cartesian-grid and arbitrary-coordinate evaluations.
We then demonstrate targeted spectral sampling in which the evaluated
frequency coordinates are concentrated toward regions containing
significant spectral weight.
Finally, using a coupled electronic dimer, we construct a hierarchy of
four models that separately isolates the effects of the RWA,
molecular evolution during the pulse envelopes, and finite spectral
bandwidth.
The resulting regime map shows that the total error of the
impulsive-RWA approximation is nonmonotonic in pulse duration and is
smallest in an intermediate regime.
For ultrashort pulses, the dominant
nonrephasing beyond-RWA correction is traced to a coupling-enabled
molecular-interaction sequence
$\mathbf{m}=(+,+,-)$ that is removed by the conventional RWA.
Together, these results establish PULSUS as a direct
frequency-domain computational tool and as a framework for diagnosing
which approximations control a given multidimensional spectrum.

%% file: Theory.tex
\section{Theory and Method}
\label{sec:theory}
\begin{figure*}[t]
    \centering
    \includegraphics[
        width=\textwidth
    ]{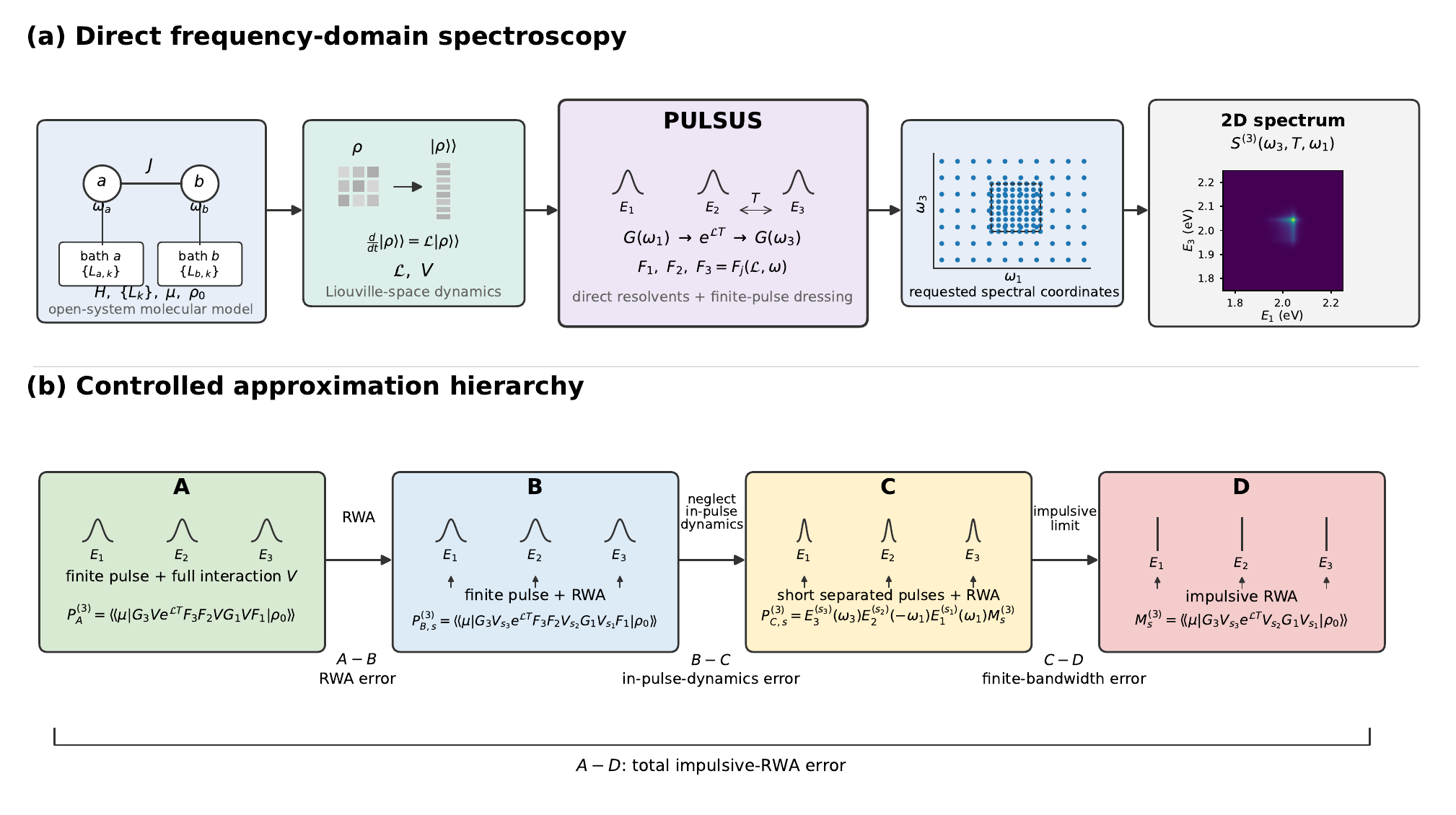}

    \caption{
    Overview of the PULSUS direct frequency-domain formulation and the
    controlled approximation hierarchy.
    (a) An open-system molecular model specified by
    $H$, $\{L_k\}$, $\mu$, and $\rho_0$ is represented in Liouville
    space and evaluated directly at requested spectral coordinates using
    Liouvillian resolvents and finite-pulse dressing operators.
    The representative spectrum shown is obtained from the full
    finite-pulse calculation.
    The plotted energy coordinates are related to the angular-frequency
    variables by $E_j=\hbar\omega_j$.
    (b) Hierarchy of models used to separate the physical approximations.
    Model A retains finite-width pulses and the complete interaction
    superoperator $V$.
    Model B applies the rotating-wave approximation while retaining the
    finite-pulse dressing.
    Model C additionally neglects molecular evolution across the pulse
    envelopes, reducing the pulse operators to scalar spectral factors.
    Model D is the impulsive RWA limit.
    Consequently, $A-B$, $B-C$, and $C-D$ isolate RWA,
    in-pulse-dynamics, and finite-bandwidth errors, respectively, while
    $A-D$ measures the total impulsive-RWA error.
    }
    \label{fig:method_workflow}
\end{figure*}

The PULSUS formulation combines open-system Liouville-space dynamics,
direct frequency-domain resolvents, and finite-pulse operator dressing.
The overall construction is summarized in Fig.~\ref{fig:method_workflow}.
We first introduce the Liouville-space representation and the direct
third-order frequency-domain response, then define the impulsive and
rotating-wave limits, and finally construct a controlled hierarchy of
finite-pulse approximations.
\subsection{Liouville-Space Formulation}
\label{sec:liouville}

We consider a finite-dimensional open quantum system described by a
density operator $\rho$ and a time-independent system Hamiltonian $H$.
Liouville-space descriptions of dissipative multidimensional
spectroscopy provide a standard framework for incorporating relaxation
and dephasing into nonlinear response calculations
\cite{
Abramavicius2009CoherentMultidimensional,
SandaMukamel2008StochasticLiouville}.
In the absence of an external field, the dynamics are assumed to be
Markovian and are written in Lindblad form,
\begin{equation}
\frac{d\rho}{dt}
=
-\frac{i}{\hbar}[H,\rho]
+
\sum_k
\left(
L_k \rho L_k^\dagger
-\frac{1}{2}
\left\{
L_k^\dagger L_k,\rho
\right\}
\right),
\label{eq:lindblad}
\end{equation}
where $L_k$ are jump operators describing dissipative and dephasing
processes.

To express the dynamics as an ordinary linear equation, we vectorize
the density operator using column-stacking. For a basis
$\{\ket{i}\}$,
\begin{equation}
\rho
=
\sum_{ij}
\rho_{ij}
\ket{i}\bra{j},
\end{equation}
we define the corresponding Liouville-space vector
\begin{equation}
|\rho\rangle\rangle
\equiv
\operatorname{vec}(\rho).
\label{eq:vectorization}
\end{equation}
The column-stacking convention obeys
\begin{equation}
\operatorname{vec}(A\rho B)
=
\left(
B^{T}\otimes A
\right)
|\rho\rangle\rangle .
\label{eq:vec_identity}
\end{equation}
The field-free equation of motion can therefore be written as
\begin{equation}
\frac{d}{dt}
|\rho(t)\rangle\rangle
=
\mathcal{L}
|\rho(t)\rangle\rangle ,
\label{eq:liouville_eom}
\end{equation}
with Liouvillian
\begin{align}
\mathcal{L}
={}&
-\frac{i}{\hbar}
\left(
I\otimes H
-
H^{T}\otimes I
\right)
\nonumber\\
&+
\sum_k
\left[
L_k^{*}\otimes L_k
-\frac{1}{2}
I\otimes L_k^\dagger L_k
-\frac{1}{2}
\left(
L_k^\dagger L_k
\right)^{T}\otimes I
\right].
\label{eq:liouvillian}
\end{align}
For a time-independent Liouvillian, the free evolution is simply
\begin{equation}
|\rho(t)\rangle\rangle
=
e^{\mathcal{L}t}
|\rho(0)\rangle\rangle .
\label{eq:free_liouville_propagation}
\end{equation}

Expectation values are represented through the Hilbert--Schmidt inner
product,
\begin{equation}
\langle\langle A|B\rangle\rangle
=
\operatorname{Tr}
\left(
A^\dagger B
\right),
\label{eq:hilbert_schmidt}
\end{equation}
so that the expectation value of an observable $A$ is
\begin{equation}
\langle A(t)\rangle
=
\langle\langle A|\rho(t)\rangle\rangle .
\label{eq:observable_liouville}
\end{equation}

Within the electric-dipole approximation, interaction with a classical
electric field $E(t)$ is described by
\begin{equation}
H_{\mathrm{int}}(t)
=
-\mu E(t),
\label{eq:dipole_interaction}
\end{equation}
where $\mu$ is the molecular dipole operator. We define the associated
light--matter interaction superoperator
\begin{equation}
V
=
\frac{i}{\hbar}
\left(
I\otimes\mu
-
\mu^{T}\otimes I
\right),
\label{eq:interaction_superoperator}
\end{equation}
such that the complete driven equation of motion becomes
\begin{equation}
\frac{d}{dt}
|\rho(t)\rangle\rangle
=
\left[
\mathcal{L}
+
E(t)V
\right]
|\rho(t)\rangle\rangle .
\label{eq:driven_liouville}
\end{equation}
This separation between field-free evolution generated by
$\mathcal{L}$ and light--matter interactions generated by $V$ provides
the starting point for constructing the nonlinear response directly
in the frequency domain.

\subsection{Direct Frequency-Domain Response}
\label{sec:direct_frequency}

Starting from Eq.~\eqref{eq:driven_liouville}, we treat the
light--matter interaction perturbatively. For three chronologically
ordered interactions, the third-order molecular response can be written
in Liouville space as
\begin{equation}
R^{(3)}(t_3,T,t_1)
=
\langle\!\langle \mu |
e^{\mathcal{L}t_3}
V
e^{\mathcal{L}T}
V
e^{\mathcal{L}t_1}
V
|\rho_0\rangle\!\rangle ,
\label{eq:third_order_time}
\end{equation}
where $t_1$ and $t_3$ are the first and third coherence intervals,
respectively, and $T$ is the waiting time. Causality requires
$t_1,t_3\geq 0$.

In a conventional time-domain calculation,
Eq.~\eqref{eq:third_order_time} is sampled over a grid of coherence
times and Fourier transformed with respect to $t_1$ and $t_3$.
Instead, we perform these transforms analytically at the level of the
Liouvillian propagator. Using the Fourier convention
\begin{equation}
\int_0^\infty
dt\,
e^{i\omega t}
e^{\mathcal{L}t},
\end{equation}
we define the causal Liouvillian resolvent
\begin{equation}
G(\omega)
=
\left[
(0^{+}-i\omega)I-\mathcal{L}
\right]^{-1},
\label{eq:resolvent}
\end{equation}
where $0^{+}$ specifies the retarded boundary condition.
In numerical calculations, the infinitesimal $0^{+}$ is replaced by a
finite positive regularization parameter $\eta$,
\begin{equation}
G_{\eta}(\omega)
=
\left[
(\eta-i\omega)I-\mathcal{L}
\right]^{-1}.
\label{eq:regularized_resolvent}
\end{equation}
Unless an additional phenomenological broadening is intentionally
introduced through $\eta$, it is chosen sufficiently small to act only
as a numerical representation of the causal $0^{+}$ prescription.
Physical relaxation and dephasing are instead included directly through
the dissipative terms in $\mathcal{L}$.
Equivalently,
\begin{equation}
G(\omega)
=
\int_0^\infty
dt\,
e^{i\omega t}
e^{\mathcal{L}t}.
\label{eq:resolvent_integral}
\end{equation}

We then define the mixed frequency--time third-order molecular response
as
\begin{equation}
M^{(3)}(\omega_3,T,\omega_1)
=
\int_0^\infty dt_3
\int_0^\infty dt_1\,
e^{i\omega_3 t_3}
e^{i\omega_1 t_1}
R^{(3)}(t_3,T,t_1).
\label{eq:mixed_response_definition}
\end{equation}
Substituting Eq.~\eqref{eq:third_order_time} and applying
Eq.~\eqref{eq:resolvent_integral} to the $t_1$ and $t_3$ propagators
gives
\begin{equation}
\boxed{
M^{(3)}(\omega_3,T,\omega_1)
=
\langle\!\langle \mu |
G(\omega_3)
V
e^{\mathcal{L}T}
V
G(\omega_1)
V
|\rho_0\rangle\!\rangle
}
\label{eq:direct_response}
\end{equation}
which is the central direct frequency-domain molecular response used
throughout this work.

Equation~\eqref{eq:direct_response} eliminates the need to propagate
the system over a two-dimensional $(t_1,t_3)$ grid and subsequently
Fourier transform the result. Instead, the response is evaluated
directly at the requested spectral coordinates
$(\omega_1,\omega_3)$. The waiting-time dynamics remain explicitly
contained in $e^{\mathcal{L}T}$, so that relaxation, population
transfer, and coherent evolution during $T$ are retained.

In numerical calculations, the action of the resolvent need not be
obtained by explicitly constructing the matrix inverse in
Eq.~\eqref{eq:resolvent}. For a Liouville-space vector
$|b\rangle\rangle$, one may instead solve
\begin{equation}
\left[
(\eta-i\omega)I-\mathcal{L}
\right]
|x\rangle\rangle
=
|b\rangle\rangle ,
\label{eq:resolvent_linear_system}
\end{equation}
with
$|x\rangle\rangle=G(\omega)|b\rangle\rangle$.
The implementation and reuse of these frequency-dependent solves are
discussed in Sec.~\ref{sec:software}.

\subsection{Impulsive and Rotating-Wave Approximations}
\label{sec:impulsive_rwa}

Equation~\eqref{eq:direct_response} is written in terms of the complete
dipole interaction superoperator $V$ and therefore contains all
excitation-raising and excitation-lowering interaction sequences
generated by the dipole commutator. In the impulsive limit, the three
field interactions are assumed to occur at sharply defined times, so
that the temporal pulse envelopes are not resolved explicitly.

To recover the conventional rotating-wave description of phase-matched
multidimensional signals, we decompose the molecular dipole operator as
\begin{equation}
\mu
=
\mu_{\uparrow}
+
\mu_{\downarrow},
\qquad
\mu_{\downarrow}
=
\mu_{\uparrow}^{\dagger},
\label{eq:dipole_decomposition}
\end{equation}
where $\mu_{\uparrow}$ and $\mu_{\downarrow}$ increase and decrease the
excitation number, respectively. The corresponding interaction
superoperators are
\begin{equation}
V_{\uparrow}
=
\frac{i}{\hbar}
\left(
I\otimes\mu_{\uparrow}
-
\mu_{\uparrow}^{T}\otimes I
\right),
\label{eq:V_up}
\end{equation}
and
\begin{equation}
V_{\downarrow}
=
\frac{i}{\hbar}
\left(
I\otimes\mu_{\downarrow}
-
\mu_{\downarrow}^{T}\otimes I
\right).
\label{eq:V_down}
\end{equation}
Thus,
\begin{equation}
V
=
V_{\uparrow}
+
V_{\downarrow}.
\label{eq:V_decomposition}
\end{equation}

For compactness, we introduce the notation
\begin{equation}
V_{+}
\equiv
V_{\uparrow},
\qquad
V_{-}
\equiv
V_{\downarrow}.
\label{eq:V_pm}
\end{equation}
Within the rotating-wave approximation (RWA), the molecular interaction
sequence is restricted to the resonant raising- and lowering-frequency
components of the dipole interaction. We label this sequence by
$\mathbf{s}=(s_1,s_2,s_3)$, where the entries are ordered chronologically
from the first to the third light--matter interaction and
$s_j\in\{+,-\}$. Here, $s_j=+$ denotes an interaction through
$V_{+}=V_{\uparrow}$, which raises the molecular excitation number,
whereas $s_j=-$ denotes an interaction through
$V_{-}=V_{\downarrow}$, which lowers it.

Different choices of $\mathbf{s}$ correspond to the conventional
phase-matched multidimensional pathways. In particular, the
nonrephasing and rephasing branches are associated with the sequences
given below
\cite{
Hybl1998TwoDimensional,
Abramavicius2009CoherentMultidimensional}.

For the nonrephasing (NR) branch, the RWA signature is
\begin{equation}
\mathbf{s}_{\mathrm{NR}}
=
(+,-,+),
\label{eq:NR_signature}
\end{equation}
giving the impulsive molecular response
\begin{equation}
M_{\mathrm{NR}}^{(3)}
(\omega_3,T,\omega_1)
=
\langle\!\langle\mu|
G(\omega_3)
V_{+}
e^{\mathcal{L}T}
V_{-}
G(\omega_1)
V_{+}
|\rho_0\rangle\!\rangle .
\label{eq:NR_impulsive_rwa}
\end{equation}

For the rephasing (R) branch, the RWA signature is
\begin{equation}
\mathbf{s}_{\mathrm{R}}
=
(-,+,+).
\label{eq:R_signature}
\end{equation}
Under the Fourier convention of
Eq.~\eqref{eq:mixed_response_definition}, the first coherence frequency
of the rephasing branch appears at negative frequency. Denoting this
signed frequency by $\nu_1<0$, the corresponding response is
\begin{equation}
M_{\mathrm{R}}^{(3)}
(\omega_3,T,\nu_1)
=
\langle\!\langle\mu|
G(\omega_3)
V_{+}
e^{\mathcal{L}T}
V_{+}
G(\nu_1)
V_{-}
|\rho_0\rangle\!\rangle .
\label{eq:R_impulsive_rwa}
\end{equation}

More generally, for any RWA interaction sequence $\mathbf{s}$, we write
\begin{equation}
M_{\mathbf{s}}^{(3)}
(\omega_3,T,\omega_1)
=
\langle\!\langle\mu|
G(\omega_3)
V_{s_3}
e^{\mathcal{L}T}
V_{s_2}
G(\omega_1)
V_{s_1}
|\rho_0\rangle\!\rangle .
\label{eq:general_impulsive_rwa}
\end{equation}

The complete interaction of Eq.~(24) contains molecular-interaction
sequences that are discarded by the RWA. It is useful here to
distinguish two notions that coincide in the conventional RWA
description but become separate once the complete interaction is
retained. The sign tuple $\mathbf{s}$ specifies the selected
positive- or negative-frequency component of each applied field and
therefore labels the phase-matched field-sign sector. Within the RWA,
that field-sign choice selects the corresponding sequence of molecular
operators $V_{s_j}$, as in Eq.~(30). Beyond the RWA, however, fixing
$\mathbf{s}$ does not restrict the molecular interaction to that single
raising/lowering sequence.

This distinction will be used below to define a controlled hierarchy
of approximations: the full finite-pulse response retains the complete
molecular interaction $V=V_{+}+V_{-}$ within a fixed field-sign sector,
whereas the impulsive-RWA response of Eq.~(30) constitutes the most
reduced description considered in this work.

\subsection{Finite-Pulse Liouvillian Dressing}
\label{sec:finite_pulse}

We now include the temporal envelopes and carrier frequencies of the
three laser pulses. 
The finite-pulse construction developed below additionally assumes
time-translation invariance of the unperturbed response. A sufficient
condition is that the state defining the response prior to the pulse
sequence is stationary under the field-free Liouvillian,
\begin{equation}
\mathcal{L}|\rho_0\rangle\rangle=0.
\label{eq:stationary_initial_state}
\end{equation}
Under this condition, the unperturbed response depends only on
interaction-time differences and the local pulse coordinates introduced
below. All finite-pulse calculations reported in this work use such
stationary initial states.
For pulse $j$, we introduce the local time coordinate
\begin{equation}
u_j = t-\tau_j,
\end{equation}
where $\tau_j$ is the pulse-center time. The real electric field is
decomposed into positive- and negative-frequency components,
\begin{equation}
E_j(u_j)
=
E_j^{(+)}(u_j)
+
E_j^{(-)}(u_j),
\end{equation}
with
\begin{equation}
E_j^{(s)}(u)
=
\frac{E_{0j}}{2}
\exp\left(
-\frac{u^2}{2\sigma_j^2}
\right)
e^{is\Phi_j}
e^{-is\omega_{Lj}u},
\qquad
s=\pm 1.
\label{eq:gaussian_field_component}
\end{equation}
Here, $\sigma_j$ is the temporal Gaussian width,
$\omega_{Lj}$ is the carrier frequency, and
$\Phi_j=\mathbf{k}_j\cdot\mathbf{R}+\phi_j$ contains the spatial and
carrier phases.

Using the Fourier convention of
Eq.~\eqref{eq:mixed_response_definition}, the corresponding scalar
laser spectrum is
\begin{equation}
E_j^{(s)}(\omega)
=
C_j e^{is\Phi_j}
\exp\left[
-\frac{\sigma_j^2}{2}
\left(
\omega-s\omega_{Lj}
\right)^2
\right],
\label{eq:gaussian_spectrum}
\end{equation}
where
\begin{equation}
C_j
=
\frac{E_{0j}}{2}
\sqrt{2\pi}\sigma_j.
\label{eq:pulse_prefactor}
\end{equation}

For finite-width pulses, the light--matter interactions may occur away
from the nominal pulse centers. We continue to assume that the pulses
are sufficiently separated that their chronological ordering is well
defined. When pulse envelopes overlap appreciably, additional temporal
orderings and response pathways can contribute and must in general be
treated explicitly
\cite{RoseKrich2021DiagramGenerator}. In particular, the causality factor associated with the second
and third interactions is approximated as
\begin{equation}
\Theta(T+u_3-u_2)
\simeq
1
\label{eq:separated_pulse_condition}
\end{equation}
over the region in which the pulse envelopes have appreciable weight.
This separated-pulse approximation does not neglect molecular evolution
during the pulse durations.

Previous finite-pulse formulations have shown that pulse-envelope
effects can enter the nonlinear response in ways that depend on the
molecular pathway and dynamics
\cite{
Smallwood2017FinitePulse,
Perlik2017FinitePulse,
DoGelinTan2017FinitePulse}.

Retaining that evolution leads naturally to operator-valued pulse
factors. For a field-sign signature
$\mathbf{s}=(s_1,s_2,s_3)$, we define
\begin{align}
F_1^{(s_1)}(\omega_1)
&=
\int_{-\infty}^{\infty}
du_1\,
e^{i\omega_1u_1}
E_1^{(s_1)}(u_1)
e^{\mathcal{L}u_1},
\label{eq:F1}
\\
F_2^{(s_2)}(\omega_1)
&=
\int_{-\infty}^{\infty}
du_2\,
e^{-i\omega_1u_2}
E_2^{(s_2)}(u_2)
e^{-\mathcal{L}u_2},
\label{eq:F2}
\\
F_3^{(s_3)}(\omega_3)
&=
\int_{-\infty}^{\infty}
du_3\,
e^{i\omega_3u_3}
E_3^{(s_3)}(u_3)
e^{\mathcal{L}u_3}.
\label{eq:F3}
\end{align}
Unlike the scalar spectra of Eq.~\eqref{eq:gaussian_spectrum}, the
operators $F_j$ are matrix functions of the field-free Liouvillian and
therefore act directly on populations and coherences in Liouville
space.

For Gaussian pulses, the integrations can be carried out analytically,
giving
\begin{align}
F_1^{(s)}(\omega_1)
&=
C_1 e^{is\Phi_1}
\exp\left[
\frac{\sigma_1^2}{2}
\left(
\mathcal{L}
+
i(\omega_1-s\omega_{L1})I
\right)^2
\right],
\label{eq:F1_gaussian}
\\
F_2^{(s)}(\omega_1)
&=
C_2 e^{is\Phi_2}
\exp\left[
\frac{\sigma_2^2}{2}
\left(
\mathcal{L}
+
i(\omega_1+s\omega_{L2})I
\right)^2
\right],
\label{eq:F2_gaussian}
\\
F_3^{(s)}(\omega_3)
&=
C_3 e^{is\Phi_3}
\exp\left[
\frac{\sigma_3^2}{2}
\left(
\mathcal{L}
+
i(\omega_3-s\omega_{L3})I
\right)^2
\right].
\label{eq:F3_gaussian}
\end{align}
Thus, finite-pulse effects appear as analytic Liouvillian-dependent
spectral filters. Although the three $F_j$ are functions of the same
Liouvillian and therefore commute with one another,
they do not generally commute with the light--matter interaction,
\begin{equation}
[F_j,V]\neq 0.
\label{eq:FV_noncommute}
\end{equation}
Their positions within the response therefore cannot be rearranged
arbitrarily.

Retaining both the finite-pulse operators and the complete molecular
interaction defines the most general separated-pulse model considered
here,
\begin{equation}
\boxed{
P_{A,\mathbf{s}}^{(3)}
=
\langle\!\langle\mu|
G(\omega_3)
V
e^{\mathcal{L}T}
F_3^{(s_3)}
F_2^{(s_2)}
V
G(\omega_1)
V
F_1^{(s_1)}
|\rho_0\rangle\!\rangle
}
\label{eq:model_A}
\end{equation}
where the frequency arguments of the $F_j$ are those defined in
Eqs.~\eqref{eq:F1}--\eqref{eq:F3}. We refer to
Eq.~\eqref{eq:model_A} as \emph{Model A}. The field-sign signature $\mathbf{s}=(s_1,s_2,s_3)$ selects the
phase-matched field contribution through the corresponding
positive- and negative-frequency components of the three pulses.
Importantly, this selection is distinct from the molecular
raising/lowering content of the response. In Model A, each molecular
interaction retains the complete superoperator
$V=V_{+}+V_{-}$, so a fixed field-sign sector contains the sum over
all molecular raising/lowering sequences allowed by the complete
interaction. Thus, ``full interaction'' here refers to the molecular
interaction content and does not imply a sum over different
field-sign or phase-matching sectors.

To isolate the effect of the rotating-wave approximation while
retaining the same finite-pulse dressing, we additionally define
\emph{Model B} by restricting each molecular interaction according to
the field signature,
\begin{equation}
P_{B,\mathbf{s}}^{(3)}
=
\langle\!\langle\mu|
G(\omega_3)
V_{s_3}
e^{\mathcal{L}T}
F_3^{(s_3)}
F_2^{(s_2)}
V_{s_2}
G(\omega_1)
V_{s_1}
F_1^{(s_1)}
|\rho_0\rangle\!\rangle .
\label{eq:model_B}
\end{equation}

A further reduction is obtained when the molecular evolution across the
pulse-centered time offsets can be neglected. In this limit,
\begin{align}
F_1^{(s_1)}(\omega_1)
&\rightarrow
E_1^{(s_1)}(\omega_1)I,
\\
F_2^{(s_2)}(\omega_1)
&\rightarrow
E_2^{(s_2)}(-\omega_1)I,
\\
F_3^{(s_3)}(\omega_3)
&\rightarrow
E_3^{(s_3)}(\omega_3)I.
\end{align}
The polarization then factorizes into scalar pulse spectra and the
impulsive RWA molecular response,
\begin{equation}
\boxed{
P_{C,\mathbf{s}}^{(3)}
=
E_3^{(s_3)}(\omega_3)
E_2^{(s_2)}(-\omega_1)
E_1^{(s_1)}(\omega_1)
M_{\mathbf{s}}^{(3)}
(\omega_3,T,\omega_1)
}
\label{eq:model_C}
\end{equation}
which we call \emph{Model C}. This is the short, separated-pulse RWA
description.

Finally, removing the finite spectral weighting gives the impulsive RWA
response
\begin{equation}
P_{D,\mathbf{s}}^{(3)}
\equiv
M_{\mathbf{s}}^{(3)}
(\omega_3,T,\omega_1),
\label{eq:model_D}
\end{equation}
which defines \emph{Model D} and corresponds to
Eq.~\eqref{eq:general_impulsive_rwa}.

The resulting sequence
\begin{equation}
A
\;\longrightarrow\;
B
\;\longrightarrow\;
C
\;\longrightarrow\;
D
\label{eq:approximation_hierarchy}
\end{equation}
forms a controlled approximation hierarchy. The $A\rightarrow B$
step introduces the RWA, $B\rightarrow C$ neglects the molecular
evolution occurring across the finite pulse envelopes, and
$C\rightarrow D$ removes the finite spectral bandwidth of the pulses.
This hierarchy will be used below to identify which approximation
dominates the error in different spectroscopic regimes.

\subsection{Physical Regimes and Dimensionless Parameters}
\label{sec:physical_regimes}

The approximation hierarchy of
Eq.~\eqref{eq:approximation_hierarchy} is governed by distinct
timescale comparisons. In particular, validity of the RWA and validity
of the short-pulse approximation are not equivalent requirements.

The importance of molecular evolution during a pulse may be
characterized by the dimensionless parameter
\begin{equation}
\chi_{\mathrm{dyn}}
=
\sigma\Omega_{\mathrm{dyn}},
\label{eq:chi_dyn}
\end{equation}
where $\sigma$ is a characteristic Gaussian field width and
$\Omega_{\mathrm{dyn}}$ is a characteristic molecular frequency
associated with the dynamics retained after the optical carrier has
been removed. When
\begin{equation}
\chi_{\mathrm{dyn}}\ll 1,
\end{equation}
the molecular state changes only weakly over the pulse duration, and
the operator-valued factors $F_j$ approach the scalar pulse spectra
used in Model C. Increasing $\chi_{\mathrm{dyn}}$ enhances the
importance of the Liouvillian dressing retained in Models A and B.

A separate dimensionless scale characterizes the rotating-wave
approximation. Counter-rotating terms oscillate with frequencies of
order
\begin{equation}
\Delta_{\mathrm{cr}}
\sim
\Omega+\omega_L,
\end{equation}
where $\Omega$ is an optical transition frequency and $\omega_L$ is
the laser carrier frequency. Near resonance,
$\Omega\simeq\omega_L$, giving the approximate scale
\begin{equation}
\chi_{\mathrm{RWA}}
\equiv
2\sigma\omega_L .
\label{eq:chi_rwa}
\end{equation}
Large $\chi_{\mathrm{RWA}}$ corresponds to many carrier oscillations
within the pulse envelope and favors dynamical suppression of
counter-rotating contributions, whereas sufficiently small
$\chi_{\mathrm{RWA}}$ indicates that the RWA may fail
\cite{Smallwood2017FinitePulse,Tan2026RWA}.

The conventional short-pulse RWA description therefore occupies an
intermediate timescale regime that can be summarized schematically as
\begin{equation}
\boxed{
\chi_{\mathrm{dyn}}\ll 1,
\qquad
\chi_{\mathrm{RWA}}\gg 1 .
}
\label{eq:short_rwa_window}
\end{equation}
These inequalities are intended as physical regime indicators rather
than universal numerical thresholds. The quantitative error depends on
the molecular spectrum, dipole couplings, relaxation rates, pulse
carrier frequencies, and the particular spectral region being probed.

For the coupled dimer considered below, a natural molecular dynamical
scale is the single-excitation splitting
\begin{equation}
\Delta E_{\mathrm{ex}}
=
\sqrt{
(E_a-E_b)^2+4J^2
},
\label{eq:exciton_splitting}
\end{equation}
so that
\begin{equation}
\chi_{\mathrm{dyn}}
=
\frac{
\sigma\Delta E_{\mathrm{ex}}
}{
\hbar
}.
\label{eq:chi_dyn_dimer}
\end{equation}
This quantity provides a convenient dimensionless indicator of
finite-pulse molecular evolution in the regime analysis.

To compare successive levels of the approximation hierarchy, we use
the full complex rephasing and nonrephasing spectra. For two models
$X$ and $Y$, with $X$ taken as the more complete description, we
define the combined relative discrepancy
\begin{equation}
\epsilon_{XY}^{\mathrm{comb}}
=
\frac{
\sqrt{
\left\|
S_Y^{\mathrm{NR}}-S_X^{\mathrm{NR}}
\right\|_F^2
+
\left\|
S_Y^{\mathrm{R}}-S_X^{\mathrm{R}}
\right\|_F^2
}
}{
\sqrt{
\left\|
S_X^{\mathrm{NR}}
\right\|_F^2
+
\left\|
S_X^{\mathrm{R}}
\right\|_F^2
}
},
\label{eq:combined_error}
\end{equation}
where
\begin{equation}
\|S\|_F
=
\left(
\sum_{mn}
|S_{mn}|^2
\right)^{1/2}
\label{eq:frobenius_norm}
\end{equation}
is the Frobenius norm.

Within the hierarchy,
$\epsilon_{AB}^{\mathrm{comb}}$ isolates the error introduced by the
RWA,
$\epsilon_{BC}^{\mathrm{comb}}$ measures the error associated with
neglecting in-pulse molecular dynamics,
$\epsilon_{CD}^{\mathrm{comb}}$ measures the effect of removing the
finite pulse bandwidth, and
$\epsilon_{AD}^{\mathrm{comb}}$ gives the total discrepancy of the
impulsive-RWA description relative to the full finite-pulse result.

%% file: Software.tex
\section{PULSUS Implementation}
\label{sec:software}

The formulation developed in Sec.~\ref{sec:theory} is implemented in
PULSUS, a Python package for direct frequency-domain spectroscopy of
open quantum systems. PULSUS provides both first- and third-order
response calculations; the present work focuses on the third-order
multidimensional formulation. The software separates the specification
of the molecular system, laser pulses, and requested spectral coordinates
from the numerical evaluation of the response. Both the complete
light--matter interaction and the controlled approximations introduced
above are exposed through a common Liouville-space representation.

\subsection{Computational Scope}

PULSUS is designed for calculations in which the spectral frequencies
themselves are the desired numerical coordinates, particularly when
only selected regions of a multidimensional spectrum are required or
when finite-pulse and rotating-wave approximations must be compared
within a common response formulation. The same response engine can
therefore be used for complete Cartesian grids, irregular coordinate
sets, local spectral refinement, and controlled comparisons among the
finite-pulse full-interaction, finite-pulse RWA, short-pulse RWA, and
impulsive-RWA descriptions.

The computational advantage of the direct frequency-domain formulation
depends on the requested spectral representation. For a Cartesian grid
containing $N_1$ excitation frequencies and $N_3$ detection frequencies,
the frequency-dependent Liouville-space objects are constructed only
along the two individual axes, requiring work associated with
$N_1+N_3$ distinct frequencies rather than independently rebuilding
the complete response at all $N_1N_3$ spectral coordinates. The final
Cartesian spectrum is obtained by contracting the resulting left- and
right-hand objects. For an arbitrary set of $K$ requested spectral
coordinates containing $N_1^{\mathrm{u}}$ unique excitation frequencies
and $N_3^{\mathrm{u}}$ unique detection frequencies, only those
$N_1^{\mathrm{u}}+N_3^{\mathrm{u}}$ frequency-dependent objects are
constructed, followed by $K$ pairwise contractions.

Consequently, PULSUS is expected to be most useful when the desired
observable consists of selected or irregular spectral coordinates, or
when repeated calculations are required at a restricted set of
frequencies. In these cases, the method avoids constructing a complete
delay-domain interferogram and avoids evaluating the Cartesian product
of spectral coordinates that were not requested. This advantage is
complementary to approaches such as UF$^2$, which efficiently treat
general sampled pulse envelopes through time-domain Fourier
convolution. The relative computational cost therefore depends on the
Liouville-space dimension, pulse representation, number and structure
of the requested spectral coordinates, and the degree to which
frequency-dependent objects can be reused; no general performance
ordering between the methods is implied here.

The present finite-pulse implementation is specialized to Gaussian
pulse envelopes, a time-independent field-free Liouvillian, and the
separated-pulse regime described in Sec.~\ref{sec:theory}. PULSUS
therefore complements, rather than replaces, approaches designed for
arbitrary sampled pulse shapes, strongly overlapping pulses, general
high-order pathway generation, or large-scale sparse and matrix-free
representations. Its present focus is a compact direct-frequency-domain
treatment in which finite-pulse dynamics, spectral-coordinate
selection, and the physical approximations entering the response can
be controlled explicitly.

\subsection{System and Pulse Representation}
\label{sec:software_system}

A molecular model is represented by the \texttt{SpectroscopySystem}
object.
The required inputs are the system Hamiltonian $H$ and the complete
dipole operator $\mu$, together with optional Lindblad collapse
operators and an initial density matrix.
From these quantities, PULSUS constructs the field-free Liouvillian
$\mathcal{L}$ and the complete interaction superoperator $V$ using the
column-stacking convention of Sec.~\ref{sec:theory}.
If no initial state is stored with the system, one may instead be
supplied when a response is evaluated; the stationary state of
$\mathcal{L}$ can also be obtained directly from the system object.

The API permits an initial density matrix to be supplied explicitly;
however, the separated-pulse finite-pulse formulation used in this work
assumes the stationary preparation of
Eq.~\eqref{eq:stationary_initial_state}.

Calculations employing the molecular rotating-wave approximation
additionally require the positive- and negative-frequency components
of the dipole, $\mu_{+}$ and $\mu_{-}$.
These components are supplied explicitly by the user and are converted
to the corresponding interaction superoperators $V_{+}$ and $V_{-}$.
Full-interaction calculations require only the complete dipole
operator and therefore do not require this decomposition.

Each finite-duration laser pulse is represented by a
\texttt{GaussianPulse} object specified by its carrier frequency
$\omega_L$, temporal width $\sigma$, peak field amplitude $E_0$, and
carrier phase $\Phi$.
These parameters determine both the scalar Gaussian spectra used in
the short-pulse model and the Liouvillian dressing operators
$F_j(\mathcal{L},\omega)$ used in the finite-pulse calculations.

\subsection{Direct Cartesian Spectral Evaluation}
\label{sec:software_cartesian}

For a Cartesian spectral grid, PULSUS exploits the separable dependence
of the third-order response on the excitation and detection frequencies.
Let $\{\omega_{1,i}\}_{i=1}^{N_1}$ and
$\{\omega_{3,j}\}_{j=1}^{N_3}$ denote the requested frequency axes.
For the finite-pulse full-interaction model, the excitation-frequency
dependent vectors are constructed as
\begin{equation}
|R_i\rangle\rangle
=
F_2(\mathcal{L},\omega_{1,i})
V
G(\omega_{1,i})
V
F_1(\mathcal{L},\omega_{1,i})
|\rho_0\rangle\rangle ,
\label{eq:software_right}
\end{equation}
while the detection-frequency dependent covectors are
\begin{equation}
\langle\langle L_j|
=
\langle\langle\mu|
G(\omega_{3,j})
V
e^{\mathcal{L}T}
F_3(\mathcal{L},\omega_{3,j}) .
\label{eq:software_left}
\end{equation}
The complete spectrum is then obtained from the contractions
\begin{equation}
S_{ji}^{(3)}
=
\langle\langle L_j|R_i\rangle\rangle .
\label{eq:software_factorization}
\end{equation}

This factorization avoids independently reconstructing the complete
Liouville-space pathway at every point of the two-dimensional grid.
For a Liouville-space dimension $n$, PULSUS stores the right-hand
objects as an $n\times N_1$ matrix and the left-hand objects as an
$N_3\times n$ matrix.
The full $N_3\times N_1$ spectrum is therefore produced by a final
matrix multiplication.
The waiting-time propagator $e^{\mathcal{L}T}$ is also constructed only
once for a given waiting time.

The frequency-domain resolvents are applied through
linear solves rather than by explicitly forming the inverse
matrix. For each requested frequency, PULSUS solves
\begin{equation}
[(\eta-i\omega)I-\mathcal{L}]|x\rangle\rangle
=
|b\rangle\rangle,
\label{eq:software_resolvent_solve}
\end{equation}
so that
$|x\rangle\rangle=G_{\eta}(\omega)|b\rangle\rangle$.

Here $\eta>0$ is the finite numerical realization of the
retarded $0^{+}$ prescription introduced in Sec.~\ref{sec:direct_frequency};
the values used in the calculations below are chosen sufficiently small
that the physical linewidths are determined by $\mathcal{L}$ rather
than by $\eta$.

On the detection side, the corresponding transposed linear system is
solved to construct the action of
$\langle\langle\mu|G_{\eta}(\omega_3)$ without explicitly evaluating
the full resolvent matrix.

The same grid-level organization is used for the finite-pulse RWA,
short-pulse RWA, impulsive RWA, and full-interaction impulsive models.
Consequently, the approximation hierarchy of
Sec.~\ref{sec:theory} can be evaluated on a common set of spectral
coordinates without changing the underlying grid representation.

\subsection{Arbitrary Spectral Coordinates}
\label{sec:software_points}

A complete Cartesian grid is not always the most efficient
representation of a multidimensional spectrum.
For example, spectral weight may occupy only a restricted region of the
frequency plane, or a reconstruction procedure may require an
irregular set of sampling coordinates.
PULSUS therefore also supports direct evaluation at arbitrary paired
spectral coordinates.

Given $K$ requested points
\begin{equation}
\left\{
(\omega_{1,k},\omega_{3,k})
\right\}_{k=1}^{K},
\end{equation}
the paired-coordinate interface evaluates only
\begin{equation}
S_k^{(3)}
=
\langle\langle
L(\omega_{3,k})
|
R(\omega_{1,k})
\rangle\rangle ,
\qquad
k=1,\ldots,K,
\label{eq:software_points}
\end{equation}
rather than constructing the Cartesian product of all distinct
$\omega_1$ and $\omega_3$ values.

The implementation further identifies repeated values along each
frequency coordinate.
If the requested set contains $N_1^{\mathrm{u}}$ unique excitation
frequencies and $N_3^{\mathrm{u}}$ unique detection frequencies, the
corresponding right- and left-hand Liouville-space objects are
constructed only
$N_1^{\mathrm{u}}$ and $N_3^{\mathrm{u}}$ times, respectively.
The previously constructed objects are then selected and contracted
pairwise according to the requested coordinates.
Consequently, the calculation does not evaluate spectral points that
are absent from the requested set.

This interface makes irregular masks, locally refined regions, and
targeted spectral sampling possible without modifying the underlying
response calculation.
The selection of sampling coordinates is deliberately separated from
the spectroscopy solver: PULSUS evaluates the coordinates supplied by
the user but does not require a particular adaptive-sampling algorithm.
This separation allows different sampling or reconstruction strategies
to be built on the same direct frequency-domain response engine.

\subsection{Field-Sign Sectors and Spectral Conventions}
\label{sec:software_signatures}

PULSUS retains the complete complex-valued third-order response and is
not restricted to the conventional rephasing and nonrephasing sectors.
The standard pathway labels provide the aliases
\begin{equation}
\mathrm{NR}\rightarrow(+,-,+),
\qquad
\mathrm{R}\rightarrow(-,+,+),
\end{equation}
but an arbitrary field-sign signature
\begin{equation}
(s_1,s_2,s_3),
\qquad
s_j\in\{-1,+1\},
\end{equation}
may instead be supplied explicitly. This permits all eight
third-order field-sign sectors to be evaluated within the
same finite-pulse or impulsive response framework. The
molecular-interaction decomposition used below to diagnose
non-rotating-wave contributions is distinct: the selected
field signature is held fixed while the raising- and
lowering-frequency components of the complete molecular
interaction are resolved separately.

For Cartesian calculations, spectra are returned as complex arrays
with the convention
\begin{equation}
S_{ji}
=
S^{(3)}(\omega_{3,j},T,\omega_{1,i}),
\end{equation}
so that the first array dimension corresponds to the detection
frequency and the second to the excitation frequency.
PULSUS uses signed first-coherence frequencies internally.
Thus, under the convention adopted here, the nonrephasing sector is
evaluated for $\omega_1>0$, whereas the rephasing sector is evaluated
for $\omega_1<0$.
For visualization, the latter may be displayed against the positive
spectroscopic excitation coordinate $-\omega_1$.

The numerical routines preserve the full complex response throughout
the calculation.
Real, imaginary, or magnitude spectra can therefore be selected only
at the analysis or visualization stage rather than being fixed by the
response solver itself.

%% file: Validation.tex
\section{Numerical Validation}
\label{sec:validation}

\subsection{External Finite-Pulse Benchmark}
\label{sec:validation_smallwood}

We first validate the finite-pulse implementation against an
independent analytical result.
Smallwood \textit{et al.} derived perturbative third-order solutions of
the optical Bloch equations for finite-duration Gaussian pulses,
providing an analytical benchmark for two-dimensional coherent
spectroscopy beyond the impulsive limit
\cite{Smallwood2017FinitePulse}.
The same model was subsequently used by Rose and Krich to validate the
UF$^2$ finite-pulse algorithm
\cite{RoseKrich2021UF2}.

We reproduce this three-level rephasing two-dimensional photon-echo
benchmark using Model B of Sec.~\ref{sec:theory}. The analytical
reference used here contains the conventional time-ordered rephasing
diagram set, and the UF$^2$ calculation is restricted to the same
diagrams. In PULSUS this corresponds to the RWA-selected rephasing
molecular-interaction sequence while retaining the finite-pulse
Liouvillian dressing.
The present comparison therefore validates this conventional
rephasing finite-pulse sector rather than the additional
molecular-interaction sequences retained by the complete-interaction
Model A.
The spectral coordinates are identified as
$\omega_1=\omega_\tau$ and $\omega_3=\omega_t$.
Because the two calculations employ different Fourier and response
normalizations, we denote the analytical reference spectrum by
$S_{\mathrm{ana}}$ and convert it to the PULSUS convention through
the fixed relation
\begin{equation}
S_{\mathrm{PULSUS}}
=
-i\,2\pi\,S_{\mathrm{ana}} .
\label{eq:smallwood_conversion}
\end{equation}
This factor follows entirely from the respective transform conventions
and is not fitted to the numerical result.

For the primary comparison, the waiting time is chosen as
$T=2\pi\sigma$, for which the pulses are sufficiently separated for
the factorization used in the PULSUS finite-pulse formulation.
Using the relative Frobenius discrepancy
\begin{equation}
\epsilon_{\mathrm{val}}
=
\frac{
\left\|
S_{\mathrm{num}}-S_{\mathrm{ana}}
\right\|_F
}{
\left\|
S_{\mathrm{ana}}
\right\|_F
},
\label{eq:validation_error}
\end{equation}
PULSUS agrees with the analytical spectrum to
$\epsilon_{\mathrm{val}}=7.33\times10^{-6}$, corresponding to a
relative discrepancy of $0.000733\%$. For the same benchmark, the
discretized UF$^2$ result used here differs from the analytical
spectrum by $0.343\%$. The UF$^2$ calculation samples each Gaussian
pulse with 25 points over a $6\sigma$ temporal window
($\Delta t=0.25\sigma$) and uses an 801-point coherence-delay grid
spanning $-100\sigma$ to $100\sigma$ with
$\Delta\tau=0.25\sigma$. The comparison is intended as an
implementation and convention check, rather than as a general accuracy
ranking between the two numerical methods.

Increasing the pulse separation to $T=8\sigma$ reduces the PULSUS
discrepancy to $1.70\times10^{-8}$.
The decrease is consistent with the remaining error at finite
separation arising from the separated-pulse time-ordering
approximation rather than from the Gaussian Liouvillian dressing
itself.

\begin{figure*}[t]
    \centering
    \includegraphics[
        width=\textwidth
    ]{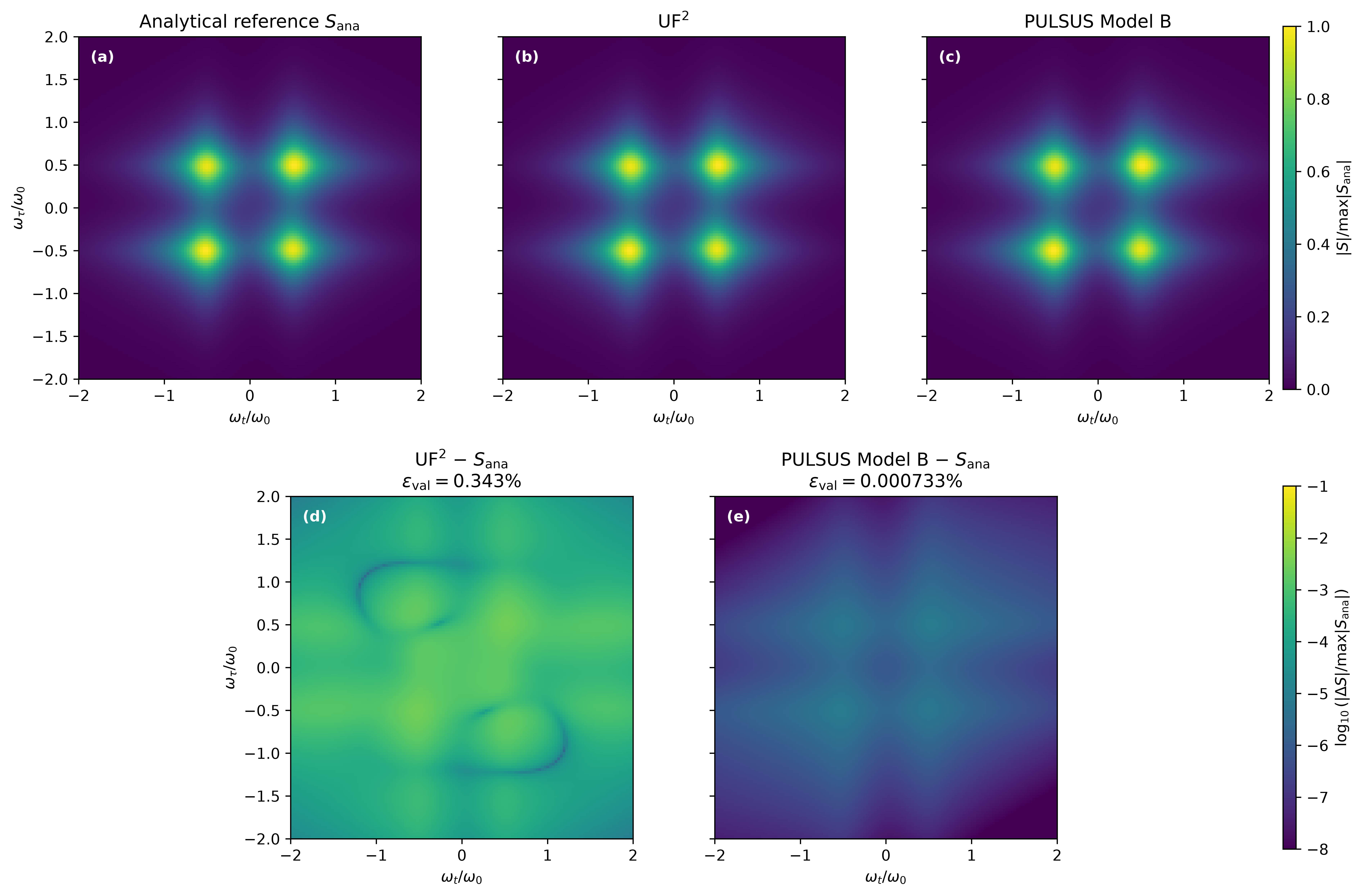}

    \caption{
    External validation of the finite-pulse implementation against the
    analytical Gaussian-pulse benchmark of Smallwood \textit{et al.}
    at waiting time $T=2\pi\sigma$.
    (a) Magnitude of the analytical reference spectrum.
    (b) Corresponding UF$^2$ result.
    (c) PULSUS Model B finite-pulse RWA result.
    The spectra in panels (a)--(c) are normalized by the maximum
    magnitude of the analytical spectrum.
    (d) Pointwise discrepancy between UF$^2$ and the analytical result,
    shown as
$\log_{10}(|\Delta S|/\max|S_{\mathrm{ana}}|)$;
    the relative Frobenius discrepancy is $0.343\%$.
    (e) Corresponding discrepancy for PULSUS Model B, with relative Frobenius discrepancy $0.000733\%$.
    The frequency axes are expressed in units of the characteristic
    frequency $\omega_0$.
    The analytical result is converted to the PULSUS Fourier convention
    using Eq.~\eqref{eq:smallwood_conversion}, with no fitted
    normalization or phase factor.
    }
    \label{fig:external_validation}
\end{figure*}

The resulting spectra and pointwise discrepancies are shown in
Fig.~\ref{fig:external_validation}.

\subsection{Full-Interaction Time-Domain Benchmark}
\label{sec:validation_fullV_ode}

The analytical Smallwood comparison above validates the conventional
finite-pulse rephasing sector represented by Model B. We therefore
perform a separate numerical benchmark of the complete-interaction
finite-pulse formulation, Model A, against an explicit time-domain
perturbative propagation.

For a fixed field-sign signature
$\mathbf{s}=(s_1,s_2,s_3)$, the reference calculation propagates the
third-order hierarchy
\begin{align}
\frac{d}{dt}|\rho^{(0)}\rangle\rangle
&=
\mathcal{L}|\rho^{(0)}\rangle\rangle ,
\\
\frac{d}{dt}|\rho^{(1)}\rangle\rangle
&=
\mathcal{L}|\rho^{(1)}\rangle\rangle
+
E_1^{(s_1)}(t)
V|\rho^{(0)}\rangle\rangle ,
\\
\frac{d}{dt}|\rho^{(2)}\rangle\rangle
&=
\mathcal{L}|\rho^{(2)}\rangle\rangle
+
E_2^{(s_2)}(t)
V|\rho^{(1)}\rangle\rangle ,
\\
\frac{d}{dt}|\rho^{(3)}\rangle\rangle
&=
\mathcal{L}|\rho^{(3)}\rangle\rangle
+
E_3^{(s_3)}(t)
V|\rho^{(2)}\rangle\rangle .
\end{align}
The complete interaction superoperator
$V=V_{+}+V_{-}$ is retained at every order, and the third-order
polarization is evaluated as
\begin{equation}
P^{(3)}(t)
=
\langle\langle\mu|\rho^{(3)}(t)\rangle\rangle .
\end{equation}
The nonrephasing and rephasing contributions are selected only through
the field signatures $(+,-,+)$ and $(-,+,+)$, respectively. The
time-domain propagation therefore does not use the analytical
Liouvillian pulse-dressing operators $F_j$ employed by Model A.

The resulting polarization is Fourier transformed using the same
frequency convention and convergence parameter as the direct
frequency-domain calculation.
The spectral-shape discrepancy is evaluated after removal of a single
global complex scale factor,
\begin{equation}
\epsilon_{\mathrm{FD/ODE}}
=
\frac{
\left\|
S_{\mathrm{ODE}}
-
\alpha S_{\mathrm{FD}}
\right\|_F
}{
\left\|
S_{\mathrm{ODE}}
\right\|_F
},
\end{equation}
where $\alpha$ minimizes the numerator.
For $\sigma=1.25$ and $T=20$, corresponding to a pulse-center
separation of $T/\sigma=16$ and therefore a well-separated-pulse
regime, and with $\eta=0.02$ in the dimensionless units of this
benchmark, increasing the Fourier window to $t_{\max}=160$ gives relative spectral-shape
discrepancies of
\begin{equation}
\epsilon_{\mathrm{FD/ODE}}^{\mathrm{NR}}
=
0.0586\%,
\qquad
\epsilon_{\mathrm{FD/ODE}}^{\mathrm{R}}
=
0.0582\%.
\end{equation}
For the nonrephasing branch, extending the Fourier window further to
$t_{\max}=200$ reduces the discrepancy to $0.0083\%$, consistent with
the remaining difference being dominated by finite numerical
transform-window effects.

The corresponding optimal global complex scale factors differ from
unity by less than $4\times10^{-6}$, showing that the agreement is not
obtained through an appreciable fitted amplitude or phase correction.

This comparison provides an independent numerical check of the
full-interaction finite-pulse implementation and complements the
analytical Model-B benchmark above.

\subsection{Internal Consistency of Spectral Evaluation}
\label{sec:validation_internal}

We next verify that the Cartesian-grid and arbitrary-coordinate
interfaces evaluate the same underlying response.
A finite-pulse Model-A spectrum was first calculated on a Cartesian
frequency grid using \texttt{finite\_pulse\_spectrum}.
The same spectral coordinates were then supplied as paired points to
\texttt{finite\_pulse\_points}, and the resulting values were reshaped
to the original grid.

The relative discrepancy was evaluated as
\begin{equation}
\epsilon_{\mathrm{grid/pts}}
=
\frac{
\left\|
S_{\mathrm{points}}-S_{\mathrm{grid}}
\right\|_F
}{
\left\|
S_{\mathrm{grid}}
\right\|_F
}.
\label{eq:grid_points_error}
\end{equation}
For the benchmark calculation,
\begin{equation}
\epsilon_{\mathrm{grid/pts}}
=
1.14\times10^{-16},
\end{equation}
showing agreement at numerical precision.
This test confirms that arbitrary-coordinate evaluation changes only
the set of requested spectral coordinates and not the underlying
finite-pulse response calculation.

The implementation additionally reuses frequency-dependent
Liouville-space objects whenever repeated values of $\omega_1$ or
$\omega_3$ occur in the requested coordinate set.
This makes the paired-coordinate interface suitable for sparse,
structured, and locally refined sampling patterns without introducing
a separate spectroscopic approximation.

\subsection{Targeted Frequency Sampling}
\label{sec:validation_sampling}

Sparse and nonuniform sampling strategies have previously been explored
as routes to reduce the experimental or computational cost of
multidimensional spectroscopy
\cite{Almeida2012CompressedSampling,Roeding2017SparseSampling}.
Here we use a simple targeted-coordinate strategy to demonstrate the
computational flexibility of the arbitrary-point PULSUS interface,
rather than to introduce a new spectral reconstruction method. The
distinctive capability being tested is that an arbitrary set of requested
$(\omega_1,\omega_3)$ coordinates can be evaluated directly without
constructing or evaluating the enclosing Cartesian frequency grid. The
particular sampling and interpolation procedure used below is therefore
illustrative and is external to the spectroscopy solver.

The arbitrary-coordinate interface also permits the spectral sampling
pattern to be adapted to the region in which the response carries
significant weight.
To demonstrate this capability, we use the same coupled-dimer model
analyzed in Sec.~\ref{sec:regime_study} and consider the full
finite-pulse nonrephasing response (Model A) at
$J=40~\mathrm{meV}$, $T=100~\mathrm{fs}$, and a pulse intensity
FWHM of $20~\mathrm{fs}$.

A dense reference spectrum was first evaluated on a
$641\times641$ Cartesian grid spanning
$1.75$--$2.25~\mathrm{eV}$ along both frequency axes.
A coarse $21\times21$ survey was then used to identify the region
containing the dominant spectral weight.
Using a threshold of $10\%$ of the maximum response magnitude,
followed by a one-cell padding of the selected region, gives the target
window
\begin{equation}
1.925 \le E_1 \le 2.075~\mathrm{eV},
\qquad
1.925 \le E_3 \le 2.100~\mathrm{eV},
\label{eq:sampling_target_window}
\end{equation}
where $E_j=\hbar\omega_j$.

We compare three sampling strategies at approximately equal coordinate
budgets. The first is a uniform Cartesian grid spanning the complete
spectral window. The second is a targeted separable grid that also
spans the complete spectral window but allocates $70\%$ of the points
along each frequency axis to the corresponding target interval. The
third is a structured local-refinement strategy that combines the
initial $21\times21$ global survey with a uniformly spaced rectangular
grid inside the target window. Duplicate coordinates between the survey
and the refined region are removed. For the two targeted strategies,
the initial survey is included in the reported total coordinate budget.

For a common reconstruction comparison, the sampled complex responses
are interpolated onto the dense reference grid using piecewise-linear
scattered-data interpolation
(\texttt{scipy.interpolate.LinearNDInterpolator}), applied
independently to the real and imaginary parts of the response. The
reconstruction error is then quantified from the full complex spectrum as
\begin{equation}
\epsilon_{\mathrm{rec}}
=
\frac{
\left\|
S_{\mathrm{rec}}-S_{\mathrm{ref}}
\right\|_F
}{
\left\|
S_{\mathrm{ref}}
\right\|_F
}.
\label{eq:reconstruction_error}
\end{equation}
This quantity measures the error introduced by the sampling and
reconstruction procedure; it is distinct from the physical
approximation errors defined in Sec.~\ref{sec:theory}.

\begin{figure*}[t]
    \centering
    \includegraphics[
        width=\textwidth
    ]{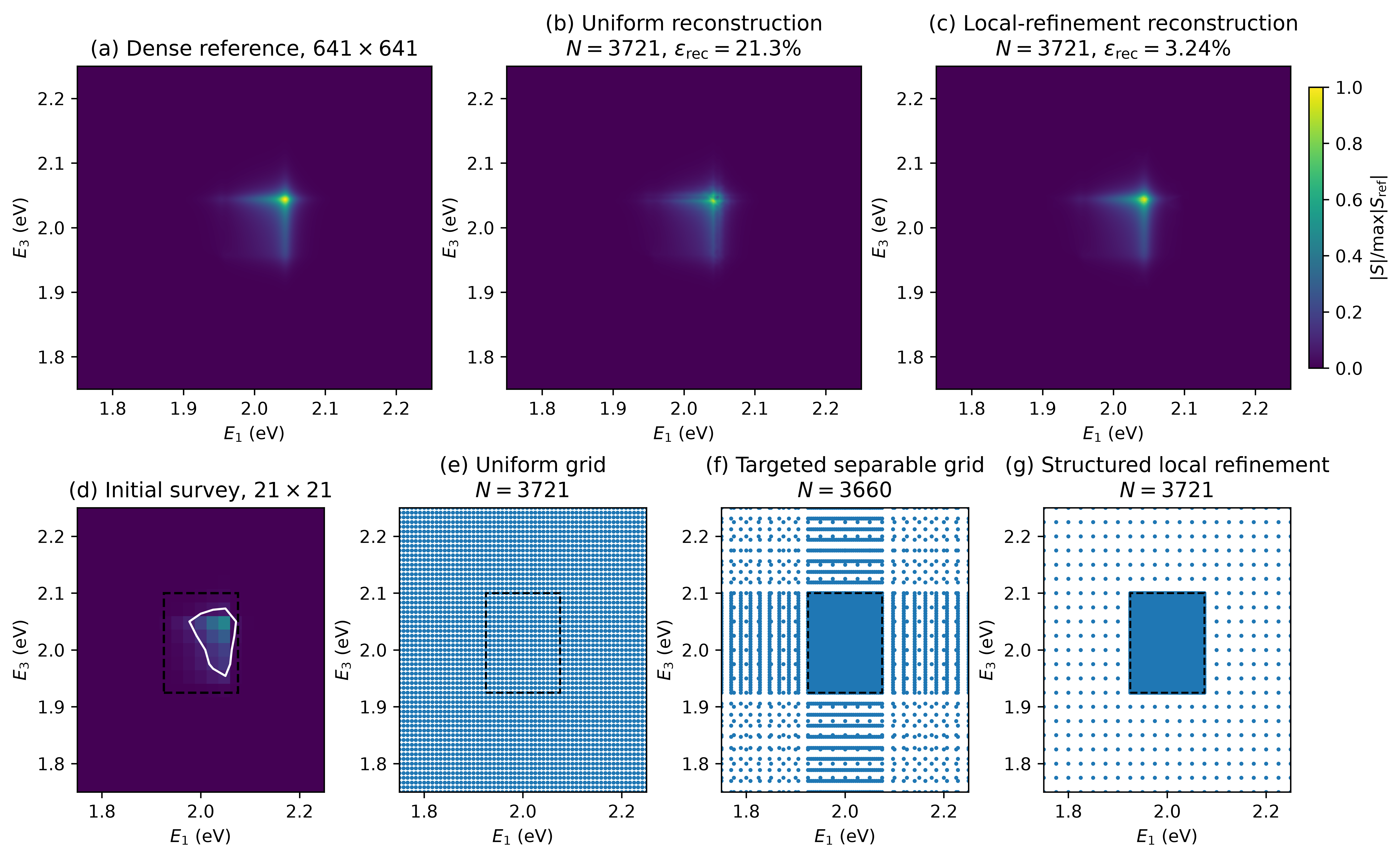}

    \caption{
    Targeted frequency sampling of the finite-pulse nonrephasing
    spectrum for the coupled-dimer benchmark.
    (a) Dense $641\times641$ reference spectrum.
    (b) Reconstruction from a uniform global grid with
    $N=3721$ sampled coordinates, giving
    $\epsilon_{\mathrm{rec}}=21.3\%$.
    (c) Reconstruction obtained from the structured local-refinement
strategy with the same coordinate budget,
giving $\epsilon_{\mathrm{rec}}=3.24\%$.
    Panels (a)--(c) show the response magnitude normalized by the
    maximum of the dense reference spectrum.
    (d) Initial $21\times21$ survey used to identify the spectrally
    active region.
    The white contour indicates the $10\%$ magnitude threshold and the
    dashed rectangle shows the padded target window defined in
    Eq.~\eqref{eq:sampling_target_window}.
    (e) Uniform global sampling grid with $N=3721$.
    (f) Targeted separable grid with $N=3660$, spanning the full spectral
window while concentrating the sampling density along the frequency
intervals identified by the initial survey.
    (g) Structured local-refinement pattern with $N=3721$, combining the
initial coarse global survey with a uniformly spaced dense rectangular
grid inside the target window.
    The sampling-coordinate selection is performed externally to the
    PULSUS response solver; each requested coordinate is evaluated
    directly using the paired-coordinate interface described in
    Sec.~\ref{sec:software_points}.
    }
    \label{fig:targeted_sampling}
\end{figure*}

At a coordinate budget of approximately $3.7\times10^3$ points,
the uniform global sampling strategy gives a reconstruction error of
$21.3\%$.
Concentrating the sampling density toward the identified spectral region
reduces the error to $6.21\%$ for the targeted separable grid and to
$3.24\%$ for the structured local-refinement strategy.
Thus, for this benchmark, redistributing a fixed sampling budget toward
the spectrally active region substantially improves reconstruction of
the full complex response.
The corresponding spectra and sampling patterns are shown in
Fig.~\ref{fig:targeted_sampling}.

\begin{figure*}[t]
    \centering
    \includegraphics[
        width=\textwidth
    ]{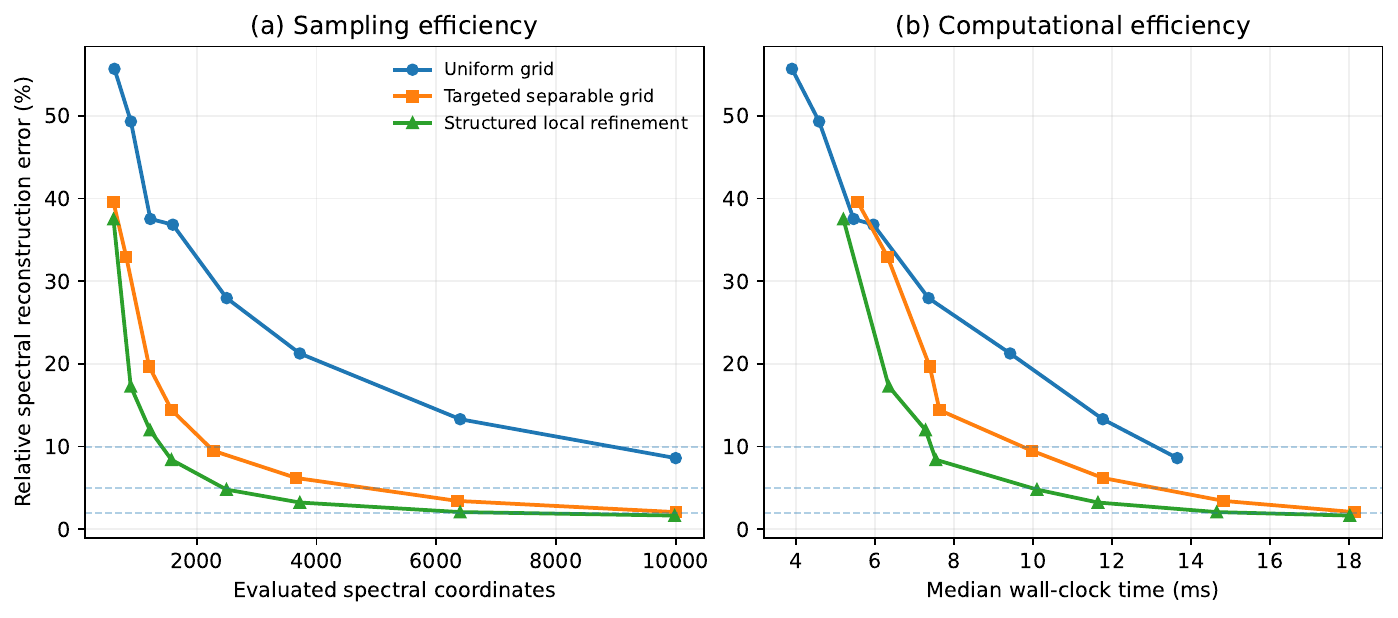}

    \caption{
    Sampling and computational efficiency of the arbitrary-coordinate
    finite-pulse calculation.
    (a) Relative complex spectral reconstruction error,
    Eq.~\eqref{eq:reconstruction_error}, as a function of the number
    of evaluated spectral coordinates for the uniform global grid,
targeted separable grid, and structured local-refinement strategy.
    (b) The same reconstruction error plotted against the median
    wall-clock time required to obtain the sampled spectrum.
    Horizontal dashed lines indicate representative reconstruction
    errors of $10\%$, $5\%$, and $2\%$.
    Near the $10\%$ level, the uniform calculation requires
    $8649$ coordinates and $13.87~\mathrm{ms}$, whereas the structured
local-refinement strategy reaches $9.10\%$ error using $1392$ coordinates
    and $7.79~\mathrm{ms}$.
    The timing for the structured strategy includes the initial coarse
    survey used to identify the target region.
    The observed efficiency gains are specific to this benchmark and
    depend on the localization of the spectral response and on the
    reconstruction strategy.
    }
    \label{fig:sampling_efficiency}
\end{figure*}

The advantage of redistributing the sampling coordinates persists over
a range of sampling budgets, as shown in
Fig.~\ref{fig:sampling_efficiency}.
For a reconstruction error of approximately $10\%$, the uniform global
grid requires $8649$ evaluated spectral coordinates and a median
wall-clock time of $13.87~\mathrm{ms}$.
The structured local-refinement strategy reaches a comparable error
($9.10\%$) using $1392$ evaluated coordinates and
$7.79~\mathrm{ms}$, including the initial survey.
For this benchmark, this corresponds to a reduction of approximately
$6.2\times$ in the number of evaluated spectral coordinates and
$1.8\times$ in wall-clock time.
The targeted separable strategy gives intermediate performance,
showing that the gain arises progressively as the coordinate budget is
concentrated toward the spectrally active region. The timing
measurements were performed on a 15-inch MacBook Air equipped with an
Apple M5 processor and 16~GB of unified memory. These factors are
specific to the present model and reconstruction procedure and are not
intended as universal performance bounds.

%% file: RegimeStudy.tex
\section{Breakdown of the Impulsive Rotating-Wave Approximation}
\label{sec:regime_study}

\subsection{Coupled Dimer Model}
\label{subsec:dimer}

To examine the breakdown of the impulsive rotating-wave description,
we consider a minimal coupled electronic dimer consisting of two
two-level chromophores, labeled $a$ and $b$.
Finite-pulse two-dimensional spectroscopy of excitonically coupled
dimers has previously been examined using Gaussian electric fields and
contrasted with the impulsive limit, including the role of pulse
overlap~\cite{Kjellberg2006Dimer}. Here we use the dimer for a
different purpose: to map the controlled
$A\rightarrow B\rightarrow C\rightarrow D$ approximation hierarchy
within the direct frequency-domain formulation.
In terms of local excitation operators $a_a$ and $a_b$, the electronic
Hamiltonian is
\begin{equation}
H
=
E_a a_a^\dagger a_a
+
E_b a_b^\dagger a_b
+
J\left(
a_a^\dagger a_b
+
a_b^\dagger a_a
\right),
\label{eq:dimer_hamiltonian}
\end{equation}
where $E_a$ and $E_b$ are the site excitation energies and $J$ is the
electronic coupling.
The local operators are restricted to the two-level site Hilbert
spaces, so that the complete model contains the ground,
singly excited, and doubly excited manifolds.

For the calculations below, the site energies are
\begin{equation}
E_a = 1.98~\mathrm{eV},
\qquad
E_b = 2.02~\mathrm{eV},
\label{eq:dimer_site_energies}
\end{equation}
giving a site-energy detuning of $40~\mathrm{meV}$.
Within the singly excited manifold, the excitonic splitting is
\begin{equation}
\Delta E_{\mathrm{ex}}
=
\sqrt{
(E_a-E_b)^2+4J^2
},
\label{eq:dimer_splitting}
\end{equation}
which gives
$\Delta E_{\mathrm{ex}}=89.4~\mathrm{meV}$
for the representative value $J=40~\mathrm{meV}$.

The dipole operator is taken to couple each site with equal transition
strength,
\begin{equation}
\mu
=
\mu_0
\left(
a_a+a_a^\dagger
+
a_b+a_b^\dagger
\right),
\label{eq:dimer_dipole}
\end{equation}
where the overall factor $\mu_0$ sets only the signal scale and is
therefore taken as unity.
For calculations employing the RWA, the corresponding
positive- and negative-frequency components are obtained directly from
the excitation-raising and excitation-lowering terms in
Eq.~\eqref{eq:dimer_dipole}.

Dissipation is described by independent local zero-temperature
Lindblad relaxation channels with jump operators

\begin{equation}
L_a=\sqrt{\gamma}\,a_a,
\qquad
L_b=\sqrt{\gamma}\,a_b,
\end{equation}

and a common relaxation rate

\begin{equation}
\gamma = 0.020~\mathrm{fs}^{-1}.
\end{equation}

Because the thermal occupation is taken to be zero, no upward thermal
jump operators are included. The initial state is the electronic
ground state.

All three laser pulses are taken to be identical, transform-limited
Gaussians centered at
\begin{equation}
\hbar\omega_L = 2.00~\mathrm{eV},
\end{equation}
with zero carrier phase.
We specify the pulse duration by its intensity full width at half
maximum,
\begin{equation}
\tau_{\mathrm{FWHM}}
=
2\sqrt{\ln 2}\,\sigma ,
\label{eq:intensity_fwhm}
\end{equation}
where $\sigma$ is the Gaussian field-envelope width introduced in
Sec.~\ref{sec:theory}.
For comparisons across pulse duration, the peak field amplitude is
chosen as
\begin{equation}
E_0
=
\frac{2}{\sqrt{2\pi}\,\sigma},
\label{eq:dimer_pulse_normalization}
\end{equation}
so that the frequency-independent Gaussian prefactor defined in
Eq.~\eqref{eq:pulse_prefactor} satisfies $C_j=1$ for each of the
three pulses at every pulse duration considered.
This normalization removes the trivial overall
$\sigma$-dependent prefactor associated with changing the pulse
duration, while retaining the frequency-dependent Gaussian spectral
weighting of Eq.~\eqref{eq:gaussian_spectrum}.

The causal resolvent regularization is set to
\begin{equation}
\eta = 10^{-10}~\mathrm{fs}^{-1},
\end{equation}
which is negligible relative to the relaxation rates and spectral
energy scales considered here. It therefore serves only as a numerical
representation of the retarded boundary condition; the physical
linewidths are generated by the Lindblad dynamics contained in
$\mathcal{L}$.

The waiting time is fixed at
\begin{equation}
T=100~\mathrm{fs}.
\end{equation}

To map the approximation regimes, we vary the electronic coupling over
\begin{equation}
0\le J\le100~\mathrm{meV}
\end{equation}
and the pulse intensity FWHM over
\begin{equation}
0.1\le\tau_{\mathrm{FWHM}}\le30~\mathrm{fs}.
\end{equation}
For each point in this parameter space, Models A--D are
evaluated on the same $161\times161$ spectral grid spanning
$1.75\leq E_1,E_3\leq2.25$~eV, corresponding to an energy
spacing of $3.125$~meV along each axis. For the rephasing
branch, the corresponding internal first-coherence frequencies
are negative, as described in Sec.~\ref{sec:software_signatures}. The models are then
compared using the combined rephasing and nonrephasing error
metric defined in Sec.~\ref{sec:theory}.

\subsection{Finite-Pulse Dynamical Effects}
\label{subsec:pulse_dynamics}

The $B-C$ comparison isolates the error introduced by neglecting
molecular evolution during the finite duration of the laser pulses.
Both models employ the rotating-wave approximation and retain the
finite spectral bandwidth of the Gaussian pulses, but they differ in
the treatment of molecular dynamics across the pulse envelopes.
Model B retains the Liouvillian dressing operators
$F_j(\mathcal{L},\omega)$, whereas Model C replaces them by scalar
spectral factors.

The importance of this approximation is controlled by the amount of
field-free evolution that can occur during a pulse.
For the coupled dimer, a natural dimensionless measure is
\begin{equation}
\chi_{\mathrm{dyn}}
=
\frac{\sigma\Delta E_{\mathrm{ex}}}{\hbar},
\label{eq:dimer_chi_dyn}
\end{equation}
where $\Delta E_{\mathrm{ex}}$ is given by
Eq.~\eqref{eq:dimer_splitting}.
When $\chi_{\mathrm{dyn}}\ll1$, the system evolves only weakly over the
pulse envelope and the short-pulse approximation is expected to be
accurate.
As either the pulse duration or the excitonic splitting increases,
molecular evolution during the pulse becomes progressively more
important and the difference between Models B and C grows.

This behavior is confirmed by the numerical scan over $J$ and pulse
duration.
For the parameter range considered here, the combined $B-C$
discrepancy spans from below $10^{-3}\%$ in the shortest-pulse regime
to approximately $46\%$ for the longest pulses and largest splittings.
The variation is smooth rather than associated with a sharp threshold,
so $\chi_{\mathrm{dyn}}$ should be regarded as a regime indicator rather
than as a universal predictor of a particular percentage error.

Importantly, this contribution is distinct from ordinary finite
bandwidth.
The separate $C-D$ comparison measures the additional effect of
removing the finite pulse bandwidth altogether. Although the constant
Gaussian prefactor is normalized to unity through
Eq.~\eqref{eq:dimer_pulse_normalization}, Model C still contains the
frequency-dependent spectral factors of
Eq.~\eqref{eq:gaussian_spectrum}, whereas Model D does not.
Consequently, the $C-D$ discrepancy includes the resulting
frequency-dependent suppression and redistribution of spectral
amplitudes and is not restricted to values below $100\%$.

\subsection{Non-Rotating-Wave Contributions}
\label{subsec:nonrwa}

The $A-B$ comparison isolates the effect of the rotating-wave
approximation while retaining the same finite-pulse treatment.
Model A uses the complete interaction superoperator
$V=V_{+}+V_{-}$ at each molecular interaction, whereas Model B
retains only the resonant sequence of raising- and lowering-frequency
molecular interactions associated with the selected spectroscopic
pathway.

For the nonrephasing signal, the phase-matched field signature remains
fixed at
\begin{equation}
\mathbf{s}_{\mathrm{NR}}=(+,-,+).
\end{equation}
To resolve the molecular origin of the beyond-RWA correction, we
expand the three complete interaction operators in Model A
independently. We denote a molecular-interaction sequence by
\begin{equation}
\mathbf{m}=(m_1,m_2,m_3),
\qquad
m_j\in\{-1,+1\},
\end{equation}
where $m_j=+1$ selects $V_{+}$ and $m_j=-1$ selects $V_{-}$ at the
$j$th molecular interaction. With the nonrephasing field signature
$\mathbf{s}_{\mathrm{NR}}$ held fixed, the full Model-A response may
therefore be decomposed as
\begin{equation}
S^{\mathrm{NR}}_{A}
=
\sum_{m_1,m_2,m_3=\pm1}
S^{(3),\mathrm{NR}}_{m_1m_2m_3}.
\end{equation}
Model B retains only the resonant molecular sequence
\begin{equation}
\mathbf{m}_{\mathrm{NR}}=(+,-,+).
\end{equation}
Thus, the eight terms in this decomposition are molecular-interaction
sequences evaluated for the same nonrephasing field signature, rather
than eight different phase-matched field-sign sectors.

As an internal consistency check, summing all eight
molecular-interaction sequences reconstructs the complete Model-A
spectrum with relative Frobenius discrepancies below $10^{-15}$ for
both the nonrephasing and rephasing signals. Moreover, the conventional
RWA-selected sequences, $\mathbf{m}_{\mathrm{NR}}=(+,-,+)$ and
$\mathbf{m}_{\mathrm{R}}=(-,+,+)$, reproduce the corresponding Model-B
spectra to numerical precision.

A useful qualitative measure of the expected RWA validity is the
dimensionless carrier-scale parameter
\begin{equation}
\chi_{\mathrm{RWA}}
\sim
2\sigma\omega_L ,
\label{eq:dimer_chi_rwa}
\end{equation}
introduced in Sec.~\ref{sec:theory}.
For long pulses, the rapidly oscillating counter-rotating contributions
are strongly suppressed and the RWA becomes accurate. For sufficiently
short pulses, this averaging is incomplete and counter-rotating
molecular-interaction sequences can contribute appreciably
\cite{Smallwood2017FinitePulse,Tan2026RWA}.
As with $\chi_{\mathrm{dyn}}$, however,
$\chi_{\mathrm{RWA}}$ is only a timescale indicator and does not by
itself determine the magnitude of the spectroscopic error.

The numerical decomposition shows that the beyond-RWA correction is
not distributed uniformly among the molecular-interaction sequences
discarded by the RWA. For short pulses, the largest omitted
contribution to the nonrephasing signal is the molecular sequence
\begin{equation}
\mathbf{m}=(+,+,-).
\end{equation}
The $0.10$~fs pulse considered here is deliberately used
as a diagnostic stress test of the RWA rather than as
a representative pulse duration for contemporary
two-dimensional electronic spectroscopy. Because such an ultrashort
pulse would have an optical bandwidth far broader than the limited
electronic manifold represented by the truncated dimer, the
calculation is not intended as a quantitatively complete model of a
real molecular experiment. Its purpose is to expose the
molecular-interaction content that emerges when carrier-cycle
averaging becomes ineffective and the counter-rotating contributions
are no longer strongly suppressed.

At a pulse intensity FWHM of 0.10~fs, its Frobenius norm is
approximately 37\% of the conventional RWA molecular sequence
$\mathbf{m}_{\mathrm{NR}}=(+,-,+)$ for $J=20$~meV and approximately
45\% for $J=40$~meV. In the uncoupled limit, $J=0$, this contribution
vanishes for the model considered here.

The corresponding spectra and coupling dependence are shown in
Fig.~\ref{fig:beyond_rwa_mechanism}.
The conventional $\mathbf{m}=(+,-,+)$ and discarded
$\mathbf{m}=(+,+,-)$ molecular sequences occupy similar regions of
the two-dimensional spectrum, but the latter is absent in the RWA and
grows rapidly with electronic coupling.
The short-pulse RWA breakdown therefore reflects a coupling-enabled
spectroscopic pathway rather than a uniform rescaling of the
conventional RWA spectrum.
More generally, the RWA error depends not only on the pulse duration
relative to the optical carrier period, but also on the molecular
couplings that determine how discarded molecular-interaction sequences
connect the available Liouville-space pathways.

\begin{figure*}[t]
    \centering
    \includegraphics[
        width=\textwidth
    ]{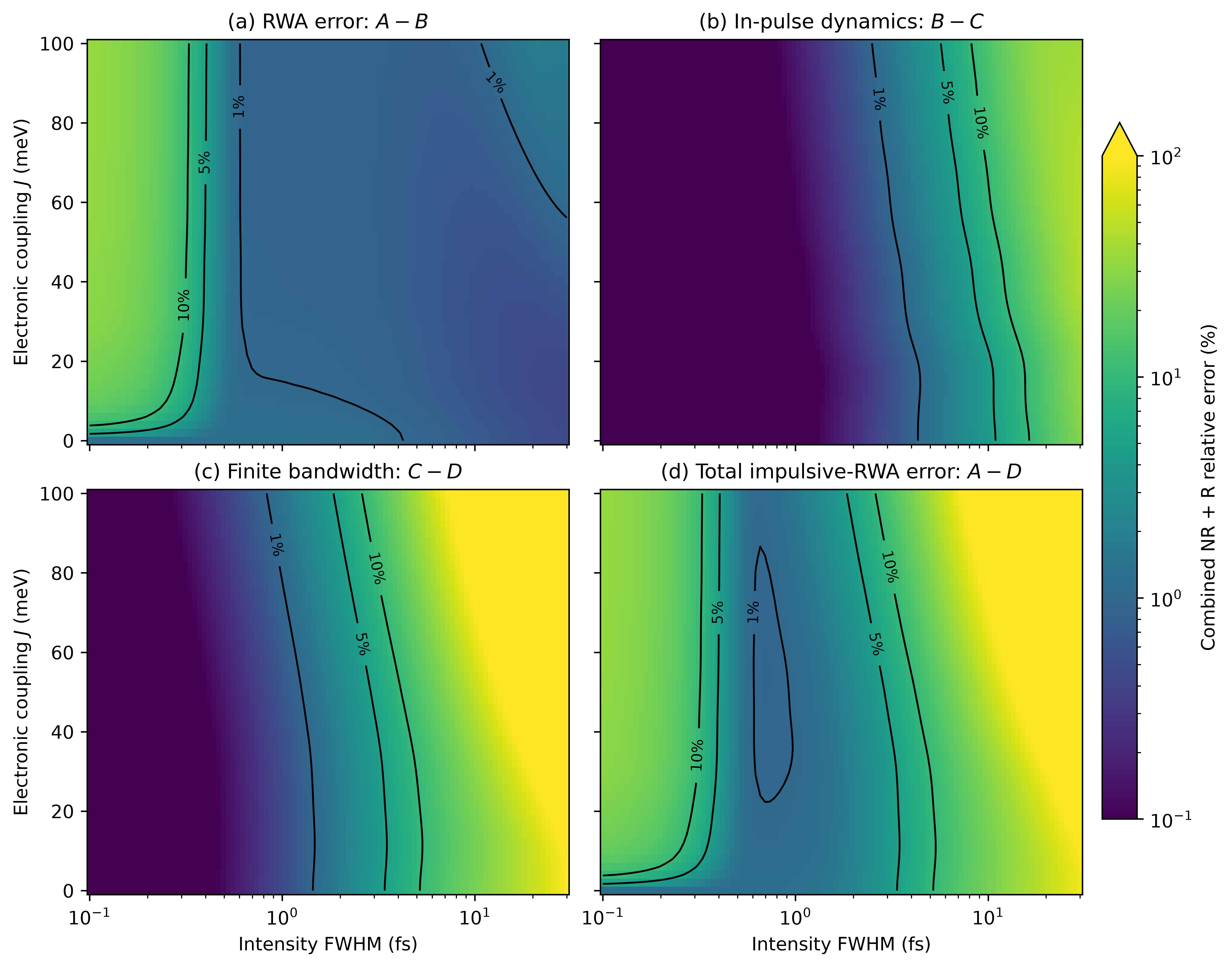}

    \caption{
    Approximation-regime map for the coupled-dimer model as a function
    of electronic coupling $J$ and pulse intensity FWHM.
    The color scale shows the combined rephasing and nonrephasing
    relative discrepancy defined in Sec.~\ref{sec:theory}, on a
    logarithmic scale.
    (a) $A-B$: error introduced by the rotating-wave approximation
    while retaining finite-pulse Liouvillian dressing.
    (b) $B-C$: error introduced by neglecting molecular evolution
    during the pulse envelopes.
    (c) $C-D$: error associated with removing the finite spectral
    bandwidth and taking the impulsive limit.
    (d) $A-D$: total discrepancy between the full finite-pulse model
    and the impulsive-RWA description.
    Black contours indicate relative errors of $1\%$, $5\%$, and
    $10\%$.
    The total error in panel (d) is nonmonotonic in pulse duration:
    the short-pulse region is limited primarily by non-RWA
    contributions, while the long-pulse region is increasingly
    affected by in-pulse molecular dynamics and finite bandwidth.
    The lowest total discrepancy in the scanned parameter range occurs
    near $J=56~\mathrm{meV}$ and
    $\tau_{\mathrm{FWHM}}=0.70~\mathrm{fs}$, where
    $\epsilon_{AD}^{\mathrm{comb}}\simeq0.93\%$.
    }
    \label{fig:regime_map}
\end{figure*}

\begin{figure*}[t]
    \centering
    \includegraphics[
        width=\textwidth
    ]{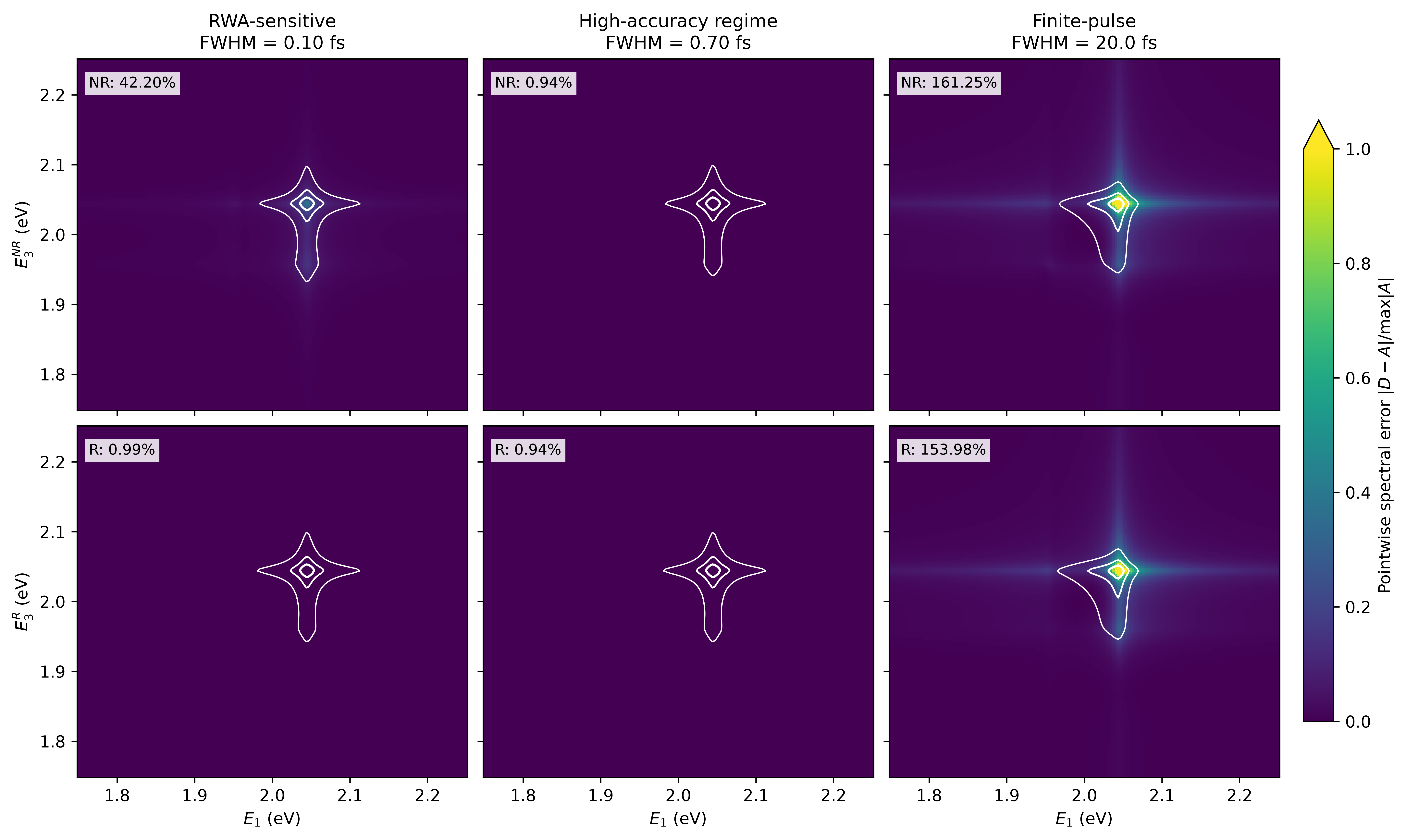}

    \caption{
    Representative pointwise spectral discrepancies between the full
    finite-pulse calculation (Model A) and the impulsive-RWA
    approximation (Model D) at $J=40~\mathrm{meV}$ and
    $T=100~\mathrm{fs}$.
    The upper row shows the nonrephasing response and the lower row
    the rephasing response.
    Columns correspond to pulse intensity FWHM values of
    $0.10~\mathrm{fs}$, $0.70~\mathrm{fs}$, and
    $20.0~\mathrm{fs}$, representing the diagnostic ultrashort RWA-stress-test, intermediate
high-accuracy, and long-pulse finite-pulse regimes,
    respectively.
    The color scale shows the pointwise difference
$|S_D-S_A|/\max|S_A|$.
White contours show the normalized magnitude of the full Model-A
spectrum, $|S_A|/\max|S_A|$, at levels of $0.10$, $0.30$, and
$0.60$, indicating where the underlying spectral response is
concentrated.
The labels in each panel give the corresponding full-complex
relative spectral discrepancy.
    At $0.10~\mathrm{fs}$ the NR response is strongly RWA-sensitive
    while the R response remains accurate; at $0.70~\mathrm{fs}$
    both pathways agree with Model A to approximately $1\%$; and at
    $20~\mathrm{fs}$ both pathways exhibit large finite-pulse
    corrections.
    }
    \label{fig:representative_spectra}
\end{figure*}

\begin{figure*}[t]
    \centering
    \includegraphics[
        width=\textwidth
    ]{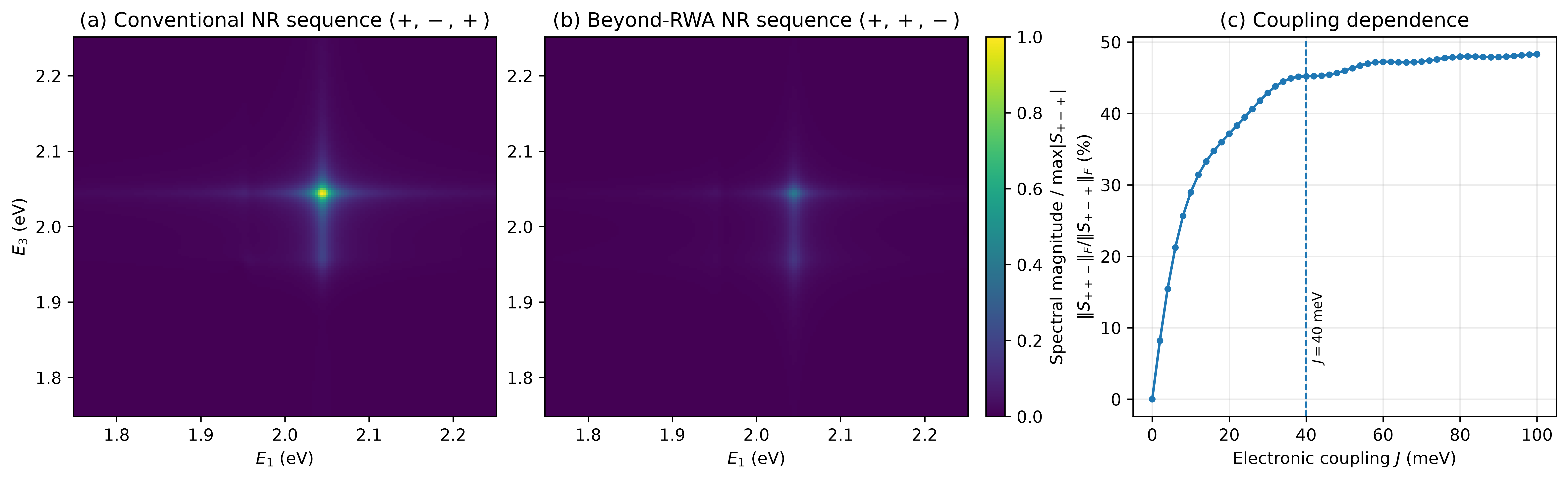}

    \caption{
    Dominant beyond-RWA molecular-interaction contribution to the
short-pulse nonrephasing response. (a) Magnitude of the conventional
RWA molecular sequence $\mathbf{m}=(+,-,+)$. (b) Magnitude of the
dominant discarded molecular sequence $\mathbf{m}=(+,+,-)$ under
the same conditions. In both panels, the nonrephasing field signature
is held fixed at $\mathbf{s}_{\mathrm{NR}}=(+,-,+)$. Both spectra are
normalized by the maximum magnitude of the conventional
$\mathbf{m}=(+,-,+)$ contribution. (c) Coupling dependence of the
Frobenius-norm ratio
$\|S_{++-}\|_F/\|S_{+-+}\|_F$
for an intensity FWHM of 0.10~fs and $T=100$~fs. The contribution
vanishes at $J=0$ and increases rapidly with electronic coupling,
reaching approximately 45\% at $J=40$~meV, indicated by the
vertical dashed line. The result identifies the
$\mathbf{m}=(+,+,-)$ molecular sequence as the dominant
coupling-enabled beyond-RWA contribution in the present
nonrephasing benchmark.
    }
    \label{fig:beyond_rwa_mechanism}
\end{figure*}

\subsection{Spectroscopic Regime Map}
\label{subsec:regime_map}

The separate comparisons above can be combined into a single map of
the approximation hierarchy.
For each electronic coupling $J$ and pulse duration
$\tau_{\mathrm{FWHM}}$, we evaluate Models A--D and compute the
combined rephasing and nonrephasing discrepancy defined in
Sec.~\ref{sec:theory}.
The four comparisons
\begin{equation}
A-B,\qquad
B-C,\qquad
C-D,\qquad
A-D
\end{equation}
therefore quantify, respectively, the RWA error, the error from
neglecting in-pulse molecular dynamics, the finite-bandwidth
contribution, and the total error of the impulsive-RWA description.

Across the parameter range considered here, the dominant source of
error changes with pulse duration.
For extremely short pulses, the finite-bandwidth correction
is small, but the rotating-wave approximation can fail
because nonresonant molecular-interaction sequences are no longer
efficiently averaged out.
At intermediate pulse durations, both the RWA and in-pulse dynamical
errors can be simultaneously small, producing a region in which the
impulsive-RWA description closely reproduces the full finite-pulse
response.
For longer pulses, molecular evolution during the pulse envelopes and
the finite spectral bandwidth become increasingly important.

The total $A-D$ discrepancy therefore exhibits a nonmonotonic
dependence on pulse duration.
For the present dimer, the smallest combined discrepancy found in the
two-dimensional scan is approximately
\begin{equation}
\epsilon_{AD}^{\mathrm{comb}}
\simeq 0.93\%,
\end{equation}
occurring near
\begin{equation}
J\simeq56~\mathrm{meV},
\qquad
\tau_{\mathrm{FWHM}}\simeq0.70~\mathrm{fs}.
\end{equation}
This minimum lies between the short-pulse regime dominated by
non-RWA contributions and the long-pulse regime dominated by
finite-pulse molecular dynamics and bandwidth.

The resulting regime map is summarized in
Fig.~\ref{fig:regime_map}.

Contours at $1\%$, $5\%$, and $10\%$ provide convenient visual guides
to the accuracy of each approximation.
These boundaries are specific to the present molecular model and
error metric; they should not be interpreted as universal thresholds.

The low-error region in Fig.~\ref{fig:regime_map}(d) forms a narrow
intermediate-pulse-duration window rather than extending monotonically
toward the impulsive limit.
At shorter durations the RWA error rises rapidly, whereas at longer
durations the $B-C$ and $C-D$ contributions increase.
The resulting minimum therefore reflects a balance between physically
distinct approximations rather than the vanishing of a single error
source.

To visualize these regimes at the level of the spectrum,
Fig.~\ref{fig:representative_spectra} shows pointwise differences
between the full finite-pulse response (Model A) and the
impulsive-RWA result (Model D) at fixed
$J=40~\mathrm{meV}$.
Three pulse durations sample the short-pulse, intermediate, and
long-pulse regimes identified in Fig.~\ref{fig:regime_map}.

At $\tau_{\mathrm{FWHM}}=0.10$~fs, the nonrephasing response is
strongly affected, with a total relative discrepancy of 42.20\%,
whereas the rephasing response remains close to the full calculation
at 0.99\%. At the intermediate duration
$\tau_{\mathrm{FWHM}}=0.70$~fs, both pathways lie in the
high-accuracy region, with discrepancies of approximately 0.94\%.
By contrast, at $\tau_{\mathrm{FWHM}}=20$~fs, the impulsive
approximation fails strongly for both pathways, giving discrepancies
of 161.25\% and 153.98\% for NR and R, respectively.

These values are relative Frobenius discrepancies and are not bounded
above by 100\%; they compare the norm of the complex spectral
difference with the norm of the full Model-A spectrum. The large
long-pulse values therefore should not be interpreted as
probability-like percentage errors. As seen in Fig.~6, the discrepancy
extends across the spectrally active region, indicating substantial
redistribution of spectral amplitude and line shape rather than an
isolated local deviation. Because the error metric is evaluated using
the full complex spectrum, phase differences also contribute to the
reported norm, although Fig.~6 visualizes the magnitude of the
pointwise spectral difference. We do not separately decompose the
total discrepancy into amplitude and phase contributions here.

The total approximation error is therefore both nonmonotonic in pulse
duration and pathway dependent.

Taken together, Figs.~\ref{fig:regime_map} and
\ref{fig:representative_spectra} show that there is no single
``short-pulse'' limit in which all approximations improve
simultaneously.
Reducing the pulse duration suppresses in-pulse molecular evolution
and broadens the available bandwidth, but eventually exposes
counter-rotating contributions that are removed by the RWA.
The intermediate high-accuracy region therefore results from the
competition between these distinct physical limits.

%% file: Conclusion.tex
\section{Conclusions}
\label{sec:conclusions}

We have developed PULSUS, a direct frequency-domain framework for
spectroscopy of open quantum systems. The method combines
Liouville-space resolvents with operator-valued Gaussian pulse dressing,
allowing requested excitation and detection frequencies to be evaluated
directly without first constructing a multidimensional time-domain
interferogram. For third-order spectroscopy, the complete
light--matter interaction may be retained, while the finite-pulse RWA,
short-pulse RWA, and impulsive-RWA descriptions arise as controlled
reductions of the same formulation.

The finite-pulse implementation was tested through complementary
benchmarks. For the conventional rephasing sector, PULSUS reproduces
the analytical Gaussian-pulse benchmark of Smallwood \textit{et al.}
with a relative discrepancy of $7.33\times10^{-6}$ at
$T=2\pi\sigma$, with the residual decreasing as the pulse separation
is increased. The complete-interaction Model A was independently
compared with explicit time-domain perturbative propagation, yielding
spectral-shape discrepancies of $0.0586\%$ and $0.0582\%$ for the
nonrephasing and rephasing responses, respectively, at the common
converged transform window. Cartesian-grid and arbitrary-coordinate
evaluations additionally agree to numerical precision.

Treating spectral coordinates as first-class computational objects
provides a natural route to selective sampling. The principal capability
demonstrated here is not the particular reconstruction algorithm, but
the ability of the PULSUS response engine to evaluate an arbitrary
requested coordinate set directly without constructing the enclosing
Cartesian product. In the coupled-dimer benchmark, concentrating the
coordinate budget toward the spectrally active region reduced the number
of evaluated points required for approximately 10\% reconstruction error
by a factor of about 6.2, with a corresponding wall-clock reduction of
about 1.8. These numerical gains depend on the model, sampling pattern,
and reconstruction procedure, but illustrate how direct spectral
evaluation can avoid computing regions of frequency space that are not
required by the analysis.

The controlled hierarchy
$A\rightarrow B\rightarrow C\rightarrow D$ also provides a direct
diagnostic of the physical approximations entering a calculated
spectrum. For the coupled dimer, the RWA, in-pulse molecular dynamics,
and finite spectral bandwidth dominate in different regions of pulse
duration and coupling strength, producing a nonmonotonic total
impulsive-RWA error. In the ultrashort-pulse stress-test regime, the
dominant nonrephasing correction discarded by the RWA is traced to the
molecular-interaction sequence
$\mathbf{m}=(+,+,-)$, whose contribution grows with electronic
coupling. This illustrates that PULSUS can be used not only to quantify
the failure of an approximation, but also to identify the
molecular-interaction content responsible for it.

The current implementation provides first- and third-order response
calculations. The finite-pulse treatment developed here specifically
assumes a time-independent field-free Liouvillian, stationary
preparation, Gaussian pulse envelopes, and sufficiently separated pulses
so that additional temporal orderings associated with pulse overlap need
not be included. Within this
scope, PULSUS is particularly useful when selected spectral coordinates
rather than complete delay grids are the desired computational objects,
when molecular evolution during finite Gaussian pulses must be retained,
or when the validity of RWA, short-pulse, and impulsive approximations
is itself part of the spectroscopic question. It therefore complements
more general pulse-shape- and pathway-oriented frameworks by providing
a compact direct-frequency-domain route in which spectral sampling and
physical approximations can be controlled explicitly within a common
response formulation.

The practical implication of the regime map is that pulse duration
alone does not establish the validity of the impulsive-RWA
approximation. Contemporary two-dimensional electronic spectroscopy
experiments commonly employ pulses in the $\sim 10$--$20$~fs range,
including measurements using approximately 12~fs and 21~fs
pulses~\cite{Timmer2023Plasmon,RojasGatjens2023ManyExciton}.
In the present coupled-dimer benchmark, this portion of the
pulse-duration range can exhibit substantial finite-pulse corrections;
for example, at $J=40$~meV and an intensity FWHM of 20~fs, the
nonrephasing and rephasing discrepancies reach 161.25\% and 153.98\%,
respectively.

These numerical values are specific to the present model and should
not be interpreted as implying that conventional calculations for
10--20~fs experiments are generally inaccurate. Rather, the accuracy
of the impulsive-RWA description depends jointly on the pulse duration
and bandwidth, molecular energy splittings and couplings, optical
carrier frequency, relaxation and dephasing scales, and the spectral
region being probed. The relevant approximations must therefore be
tested relative to the molecular and optical timescales of the system
under study rather than inferred from pulse duration alone. PULSUS
provides such a test by placing the complete finite-pulse response and
its RWA, short-pulse, and impulsive reductions within a common
computational framework.

%% file: EndMatter.tex
\begin{acknowledgments}

This work was funded by the National Science Foundation (CHE-2404788), Robert A.\ Welch Foundation (E-1337), and the U.S. Department of Energy, Office of Science, under Award No. DE-SC0025706. ERB gratefully acknowledges funding from IVADO for a Visiting Professorship at the Institut Courtois, Universit\'e de Montr\'eal. 

\end{acknowledgments}

\section*{AUTHOR DECLARATIONS}

\subsection*{Conflict of Interest}

The authors have no conflicts to disclose.

\subsection*{Author Contributions}

A.C.-G. developed the finite-pulse Liouvillian dressing formalism,
implemented the PULSUS software, performed the numerical calculations
and validation studies, analyzed the results, prepared the figures,
and wrote the original draft of the manuscript. E.R.B. proposed the
initial use of Liouvillian Green's-function methods for the
frequency-domain formulation, supervised the project, and contributed
to the conceptual development, interpretation of the results, and
revision of the manuscript. Both authors discussed the results and
approved the final version of the manuscript.

A.C.-G. acknowledges the use of ChatGPT (OpenAI) as an assistive
tool during software development, code debugging, numerical workflow
development, and manuscript preparation. All AI-assisted material,
scientific results, interpretations, and final manuscript content were
reviewed and verified by A.C.-G.

\section*{DATA AVAILABILITY}

PULSUS is open-source software distributed under the BSD 3-Clause
license. The source code is publicly available through GitHub, and the
package can be installed from the Python Package Index as
\texttt{pulsus-spec}. The version used for the calculations reported
in this work, PULSUS v0.2.0, is archived on Zenodo under the
version-specific DOI 10.5281/zenodo.23253040. Archived releases are
also available through the Zenodo concept DOI
10.5281/zenodo.22755139.